\documentclass{nature}

\usepackage{comment}

\usepackage{amsmath}
\usepackage[dvipsnames]{xcolor}
\usepackage{soul}
\usepackage{graphicx}
\usepackage{amssymb}

\usepackage{subcaption}
\usepackage{tabularx}
\usepackage{ragged2e}

\usepackage{multirow}

\usepackage{url}

\usepackage[symbol]{footmisc}

\newcommand{\gaia}{\textit{Gaia}}
\newcommand{\chandra}{\textit{Chandra}}

\usepackage[switch, modulo]{lineno}
\modulolinenumbers[1]

\def\apjs{{ApJS}}               
\def\nat{{Nature}}              

\def\aap{{Astron. Astrophys.}}                
\def\araa{{Annu. Rev. Astron. Astrophys.}}             
\def\apj{{Astrophys. J.}}                 
\def\apjl{{Astrophys. J. Lett.}}                
\def\apjs{{Astrophys. J. Suppl.}}               
\def\mnras{{Mon. Not. R. Astron. Soc.}}             
\def\pasp{{Publ. Astron. Soc. Pacif.}} 			

\title{A white dwarf accreting planetary material determined from X-ray observations}

\author{Tim Cunningham$^{1,2}\star$,
Peter J. Wheatley$^{1,2}$, 
Pier-Emmanuel Tremblay$^{1,2}$, 
Boris T. G\"{a}nsicke$^{1,2}$, 
George W. King$^{3}$, 
Odette Toloza$^{4,5}$ \&
Dimitri Veras$^{1,2}$
}

\begin{document}

\maketitle

\begin{affiliations}
 \item Department of Physics, The University of Warwick, Coventry, CV4 7AL, UK
 \item Centre for Exoplanets and Habitability, University of Warwick, Gibbet Hill Road, Coventry CV4 7AL, UK
 \item Department of Astronomy, University of Michigan, Ann Arbor, MI 48109, USA
 \item Departamento de Física, Universidad Técnica Federico Santa María, Av.\,España 1680, Valparaíso, Chile
 \item Millennium Nucleus for Planet Formation (NPF), Valparaíso, Chile
\end{affiliations}

\begin{abstract}

The atmospheres of a large proportion of white dwarf stars are polluted by heavy elements\cite{koester14} that are expected to sink out of visible layers on short timescales\cite{paquette86a,koester09}.
This has been interpreted as a signature of ongoing accretion of debris from asteroids\cite{zuckerman07}, comets\cite{xu17}, and giant planets\cite{gaensicke19}.
This scenario is supported by the detection of debris discs\cite{zuckerman87} and transits of planetary fragments\cite{vanderburg15} around some white dwarfs.
However, photospheric metals are  only indirect evidence for ongoing accretion, and the inferred accretion rates and parent body compositions heavily depend on models of diffusion and mixing processes within the white dwarf atmosphere\cite{bauer2019,cunningham19,heinonen2020}. 
Here we report a 4.4$\sigma$ detection of X-rays from a polluted white dwarf, G29--38, using a 106\,ks exposure with the \textit{Chandra X-ray Observatory}\cite{weisskopf2000Chandrafornaturesubmission}, demonstrating directly that the star is currently accreting. 
From the measured 
X-ray luminosity,
we find an instantaneous accretion rate of
$\dot{M_{\rm X}}=1.63^{+1.29}_{-0.40}\times 10^{9}\mathrm{\,g\,s^{-1}}$. 
This is the first direct measurement of the accretion rate onto the white dwarf, which is independent of stellar atmosphere models. This rate exceeds estimates based on past studies of the photospheric abundances by more than a factor two, and implies that convective overshoot has to be accounted for in modelling the spectra of debris-accreting white dwarfs.
We measure a low plasma temperature of $kT=0.5\pm0.2\,\mathrm{keV}$, corroborating the predicted bombardment solution for white dwarfs accreting at low accretion rates\cite{kuijpers1982,woelk1993}.
Offering a new method for studying evolved planetary systems, these observations provide the opportunity to independently measure the instantaneous accretion rate of planetary material, and therefore investigate the timescale of accretion onto white dwarfs\cite{farihi2018XRay,cunningham2021}, and the evolution and replenishment of debris disks\cite{kenyon2017b}.
\end{abstract}

G29--38 is among the 100 closest white dwarfs\cite{hollands18b} and has hence been subject to detailed studies across most wavelength ranges. The detection of an infrared excess was initially interpreted as arising from a brown dwarf companion\cite{zuckerman87}. However, subsequent ultraviolet (UV) spectroscopy revealed trace metals in the hydrogen atmosphere of the white dwarf\cite{koester97}, which were  interpreted as the signature of ongoing accretion from a compact dusty debris disk that formed from the tidal disruption of an asteroid\cite{jura03}. Assuming that the white dwarf atmosphere is in an equilibrium between accretion and gravitational settling predicts\cite{xu14} an accretion rate of $6.5\times10^8\,\mathrm{g\,s^{-1}}$. Some 25--50\% of Milky Way white dwarfs are expected to be metal-polluted\cite{koester14}, and
there are now more than a thousand known metal-polluted stellar remnants\cite{coutu19},
and yet the evidence for ongoing accretion remains circumstantial, based on the modelling of atmospheric  abundances.

White dwarf accretion should be accompanied with intense heating of the infalling material, sufficient to promote cooling via X-ray emission\cite{mukai2017}.
This has been observed directly 
for white dwarfs accreting from stellar companions\cite{mukai2017,patterson85,woelk1993,kuijpers1982}, but never for a white dwarf accreting planetary debris.
An \textit{XMM-Newton} observation of G29--38 resulted in a non-detection due to the presence of a bright X-ray source nearby, 
placing
an upper limit\cite{farihi2018XRay} on the accretion rate of $\simeq10^{10}\,\mathrm{g\,s^{-1}}$. Similar upper limits from non-detections have also been derived for a handful of other metal-polluted white dwarfs\cite{farihi2018XRay}, but a detection of X-rays -- direct evidence of ongoing accretion -- has never previously been achieved. 

\begin{figure}
    \RaggedRight
    \begin{tabularx}{\columnwidth}{XX}
        \textsf{\textbf{a}} & \textsf{\textbf{b}}
    \end{tabularx}
    \centering
    \justifying
	\includegraphics[width=.49\columnwidth]{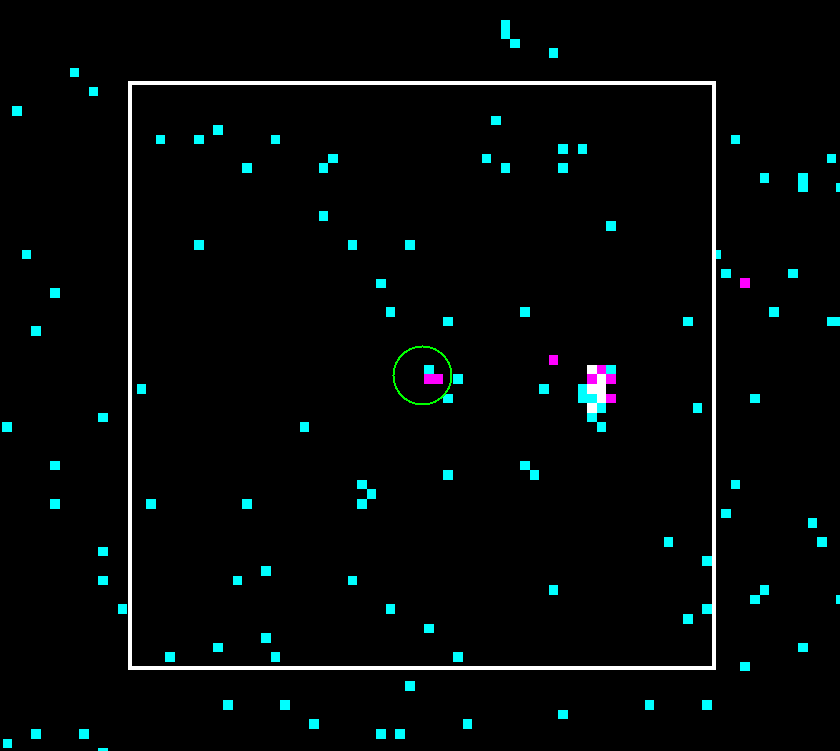}
	\hfill
	\includegraphics[width=.49\columnwidth]{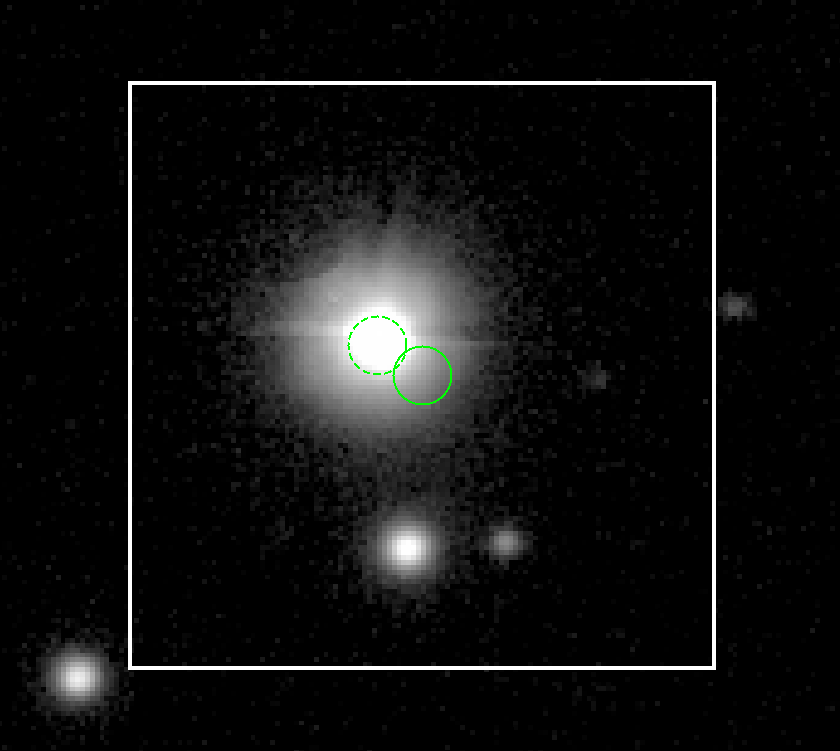}
	\caption{\textbf{X-ray and optical imaging of G29--38.} \textbf{a}, Image of the recorded events in the soft+medium (0.5--2.0\,keV) band. Black, cyan, magenta and white represent 0, 1, 2, and 3 counts, respectively. \textbf{b}, The PanSTARRS1 (PS1) $i$-band image. In both panels, the white square, centered on the target coordinates, has sides of length 30\,arcsec. The solid green circle shows 1.5\,arcsec radius around target coordinates, whilst the dashed green circle shows the target position at the PS1 epoch (J2014.89). The \textit{Chandra} observations reveal a significant X-ray source at the expected position of G29--38.}
	\label{fg:PanSTARRS-Chandra}
\end{figure}

We observed G29--38  with \textit{Chandra} ACIS-S on five occasions between 22 and 27 September 2020, with a total exposure time of 106.33\,ks (see Extended Data Table\,\ref{tab:Chandra-observations} for details). The sky location of the X-ray photons of our observations is shown in Fig.\,\ref{fg:PanSTARRS-Chandra}, along with an image at optical wavelengths.
The data reduction utilised three standard \textit{Chandra} science bands: soft (0.5--1.2\,keV), soft+medium (0.5--2.0\,keV) and broad (0.5--7.0\,keV).
Within 1\,arcsec of our target coordinates, 
we detected a total of five 
X-ray
events; 
with four in the soft band and one in the medium band. Using a Bayesian approach\cite{kraft1991} (see Methods), we found 
the 68\% confidence interval on the source count rate in the broad band to be $(2.4$--$6.6)\times10^{-5}$ counts per second
(see Extended Data Table\,\ref{tab:confidence-interval-counts} for count rates in the soft and soft+medium bands).

\begin{figure}
  \centering
  \includegraphics[angle=0,width=.7\textwidth]{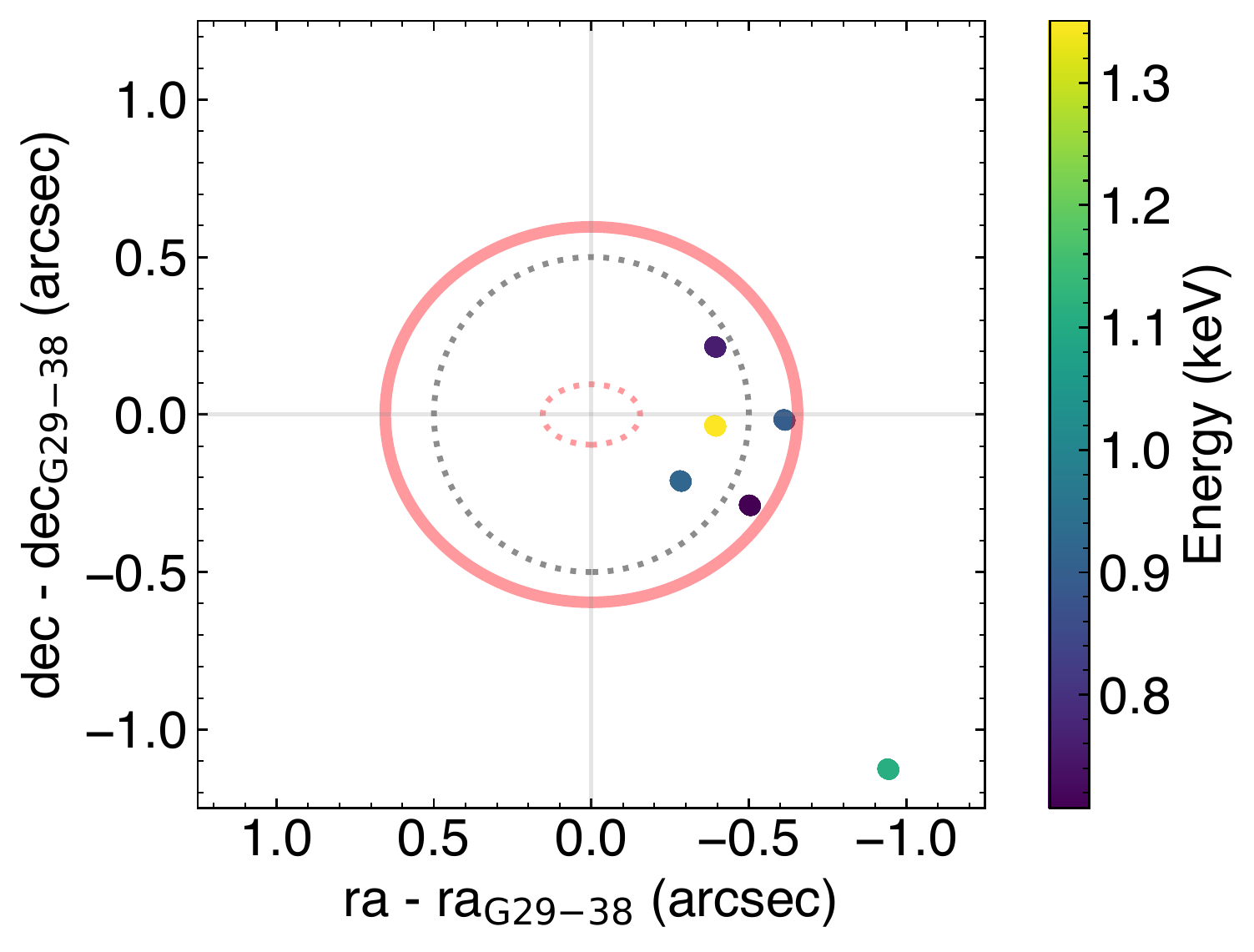}
  \caption{\textbf{Sky location of X-rays from G29--38.} Position of soft+medium band (0.5--2.0\,keV) X-ray events in the merged observations.
  The colour of each event indicates the measured energy of the event. Also shown are the 1$\sigma$ uncertainty contours on the target position. The innermost, red-dashed ellipse gives the target positional uncertainty from \textit{Gaia} EDR3. The wider, grey-dashed circle indicates the 68\% positional uncertainty of 0.5\,arcsec, arising from the astrometric accuracy of \textit{Chandra} ACIS-S (see figure 5.4 of the \chandra\ Proposers' Observatory Guide Version 23.0\footnote{https://cxc.harvard.edu/proposer/POG/pdf/MPOG.pdf}). The outer, red-solid ellipse shows the combined 68\% uncertainty interval on the position. The five source photons used in our statistical analyses are all consistent with the expected sky location of G29--38.}
  \label{fg:RA-DEC-events-uncertainty}
\end{figure}

Fig.\,\ref{fg:RA-DEC-events-uncertainty} shows the sky location of the soft+medium band events, which are fully consistent with the 1$\sigma$ uncertainty on the position from the combined \textit{Gaia} EDR3\cite{gaia2021EDR3} and \textit{Chandra} ACIS-S astrometry.
Considering the three science bands
we computed the probability that the observed events could arise by chance. We used a standard aperture photometry  approach (see Methods) with source counts of 4, 5, and 5, in the soft, soft+medium and broad bands, respectively, and a measured background 
of 
0.10, 0.21 and 0.57\,counts per 1 arcsec aperture, respectively. 
Using the Poisson distribution, we calculated the probability of 
receiving the observed source counts or higher
within 1\,arcsec, given the observed background count rate,
and found the statistical significance of the source counts in the three bands
 to be 4.62, 4.71 and 3.64$\sigma$ 
(see Extended Data Table\,\ref{tab:Poisson-source-and-background}), indicating a high-confidence detection of X-rays from G29--38. We note that accounting for the astrometric uncertainty by centring the aperture on the detected source position (Fig.\,\ref{fg:RA-DEC-events-uncertainty}), 
and using a smaller radius of 0.5\,arcsec, which is appropriate for soft sources in ACIS-S 
(see Methods),
increases the detection significance to 5.65, 5.94 and 5.06$\sigma$ in the three science bands, respectively. 

As confirmation that the observed source counts originated from the target, and not a background source, 
we utilised a source detection algorithm (see Methods) to derive a sky density of sources at the depth of the ACIS-S observations.
In the soft, soft+medium, and broad bands, this analysis resulted in 7, 25, and 32 sources, respectively, setting the background source sky density over the whole image to $\rho_{\rm sky}=3$, $10$ and $13\times10^{-5}$\,arcsec$^{-2}$. The detected sources comprise real astrophysical sources as well as any spurious detections arising from background counts. Interpreting this as the probability of a background source chance-aligning 
within 1\,arcsec of our target, we were able to 
confirm the source events originated from our target
with a statistical significance of 
4.2, 3.9 and 3.8$\sigma$.
Finally, 
as an independent confirmation,
we also performed a Monte Carlo aperture photometry analysis (see Methods) within 100\,arcsec of the target. In the soft band image, out of 100,000 test apertures, 
only 0.001\% retrieved four events, with none retrieving more than four. This allows us to rule out chance alignment with a background source at 4.4$\sigma$. 

We conclude that the four recorded events in the standard soft science band reveal a $>4\,\sigma$ detection of X-rays from G29--38, with the source detected at 4.62 or 5.65$\sigma$, depending on the aperture size, and chance alignment ruled out at 4.4$\sigma$.
The energies of the five recorded events within 1\,arcsec of our target are all in the range 0.7--1.4\,keV, with four below 1\,keV. As the sensitivity of the ACIS-S detector is greatest in the range $\approx$1--6.5\,keV, and drops off steeply towards lower energies (see Extended Data Fig.\,\ref{fg:counts-effective-area}), the detection of most photons
below 1\,keV
strongly suggests that the X-ray emission spectrum is 
very soft, and thus emitted from a relatively low-temperature plasma. 
To determine the best-fitting plasma temperature, $T_{\rm X}$, we initially adopted an optically-thin isothermal plasma model and tested two distinct sets of abundances: bulk Earth\cite{mcdonough1995} and the spectroscopically determined photospheric abundances of G29--38\cite{xu14}. 
The plasma temperatures were consistent, with the photospheric abundances corresponding to $kT_{\rm X}=0.49^{+0.17}_{-0.29}$\,keV (see Extended Data Table\,\ref{tab:confidence-interval-kT}). 

While no debris-accreting white dwarf has previously been detected at X-ray wavelengths, accretion onto white dwarfs at much higher rates is common in close white dwarf binaries.
Our measured plasma temperature of 0.5\,keV from a debris-accreting white dwarf is much lower than
for white dwarfs accreting from stellar companions, which typically accrete at rates 
many orders of magnitude higher  ($\gtrsim10^{16}\,\mathrm{g\,s^{-1}}$) and have plasma temperatures\cite{mukai2017} in the range 5--50\,keV.
This is a robust result because the sensitivity of the ACIS-S detector increases to higher energies (peaking the range 1--6.5\,keV) and emission with the same luminosity at higher temperatures would have been readily detected. Measuring a temperature an order of magnitude lower than other accreting white dwarfs points to a heating mechanism that is distinct from the strong stand-off shocks thought to heat the infalling material at higher accretion rates\cite{patterson85}.  A possible explanation is that the very low accretion rate we observe for G29-38 is not sufficient to support a stand-off shock, and the infalling material impacts directly onto the white dwarf 
surface: the previously proposed
``bombardment'' solution\cite{kuijpers1982,woelk1993}. For a white dwarf with mass $M_{\rm WD} = 0.6\,M_{\odot}$ and radius $R_{\rm WD} = 0.0129\,R_{\odot}$, the plasma temperature arising from accretion in the bombardment scenario was predicted by ref.\cite{kuijpers1982} (see their equation 9) to be $T_{\rm X}\approx0.6$\,keV, which is consistent with the measured plasma temperature from our observations.

In order to determine the X-ray flux, we also adopted a more physically-motivated spectral model 
that combines emission from an optically-thin plasma at a range of temperatures
(a cooling-flow model; see Methods
for the full set of model fits). 

The best-fit model was integrated over the \textit{Chandra} ACIS-S passband (0.3--7.0\,keV) to derive an X-ray flux of $F_{\rm X}(0.3{-}7.0\,\rm{keV})=1.97^{+1.55}_{-0.48}\times 10^{-15}\,\mathrm{erg\,s^{-1}\,cm^{-2}}$. We also perform the integration over a wider energy band to estimate the flux at unobserved wavelengths.
From the  \textit{Gaia} EDR3 parallax, the distance to G29--38 is 
$d=(17.53\pm0.01)$\,pc 
which implies a best-fit X-ray luminosity of 
$L_{\rm X}(0.3{-}7.0\,{\rm keV)}=7.24^{+5.66}_{-1.76}\times10^{25}\,\mathrm{erg\,s^{-1}}$.
This is many orders of magnitude lower than observed for white dwarfs accreting from main-sequence companions (typically $L_{\rm X}=10^{29}{-}10^{33}\rm\,erg\,s^{-1}$)\cite{mukai2017}.

\begin{figure}
	\centering
	\includegraphics[width=.7\columnwidth]{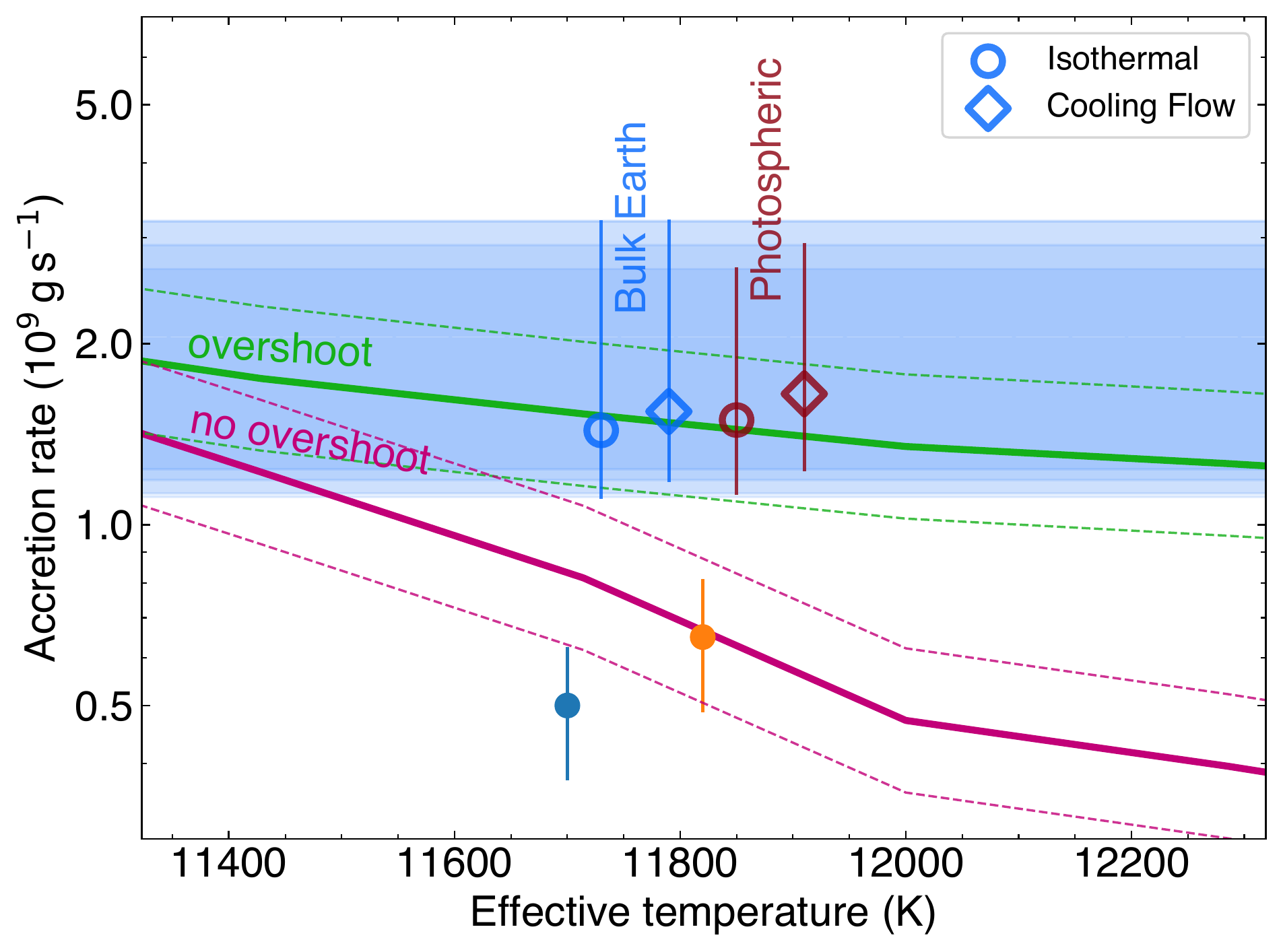}
	\caption{\textbf{Accretion rates inferred from the measured X-ray flux.} Accretion rates from the isothermal and cooling flow models are shown as open circles and diamonds, respectively. We show results for two plasma abundances; bulk Earth\cite{mcdonough1995} and photospheric\cite{xu14} 
	shown with blue and red, respectively. The filled horizontal bands show the 68\% confidence interval on the four accretion rates. The X-ray accretion rates are computed using Equation\,\eqref{eq:Mdot-Xray}.
	Using Equation\,\eqref{eq:Mdot-steady-state}, the accretion rates inferred from spectroscopic observations are shown in solid lines for a $0.6M_{\odot}$ white dwarf, assuming bulk Earth composition; such that the observed calcium abundance\cite{xu14} of $\log{\rm [Ca/H]}=-6.58 \pm0.12$ makes up 1.6\% of the accreted material. Accretion rates from models which include or omit convective overshoot mixing\cite{cunningham19} are shown in green or magenta, respectively, with 1$\sigma$ uncertainties (dashed lines) propagated from the spectroscopic abundance measurement. Also shown in solid circles are the previously published inferred accretion rates for G29--38 in blue\cite{farihi09} and orange\cite{xu14}. 
	}
	\label{fg:Mdot-1D-3D}
\end{figure}

The instantaneous accretion rate can be estimated from the X-ray luminosity using\cite{patterson85}
\begin{equation}
 \dot{M_{\rm X}} = \frac{2}{A}L_{\rm X} \frac{R_{\rm WD}}{GM_{\rm WD}}\, ,
 \label{eq:Mdot-Xray}
\end{equation}
where $R_{\rm WD}=0.0129\,R_{\odot}$ is the white dwarf radius,
$M_{\rm WD}=0.6 M_{\odot}$ is the white dwarf mass,
$G$ is the gravitational constant, $L_{\rm X}$ is the measured X-ray luminosity and the factor two accounts for 50\% of the emitted photons being directed back towards and absorbed by the star\cite{kylafis1982}. The constant $A$ quantifies the fraction of the total luminosity, from accretion, carried in the observed band (0.3--7.0\,keV). 
To calculate the X-ray accretion rate from our observations,
we take the limiting case that the total luminosity is equivalent to the observed X-ray luminosity ($A=1$).
We allow the plasma to be formed near the white dwarf surface, 
which is supported by the bombardment scenario of accreted material directly impacting the photosphere\cite{kuijpers1982}.
Under these assumptions, Fig.\,\ref{fg:Mdot-1D-3D} shows the 68\% confidence interval on the instantaneous accretion rate derived for the two plasma models (isothermal and cooling flow) and two debris abundances (bulk Earth and photospheric), across the range of uncertainty on the photospheric effective temperature of G29--38. Using Equation\,\eqref{eq:Mdot-Xray}, the bulk Earth abundances and cooling flow model, we find the best-fit instantaneous accretion rate to be 
$\dot{M_{\rm X}}(0.3{-}7.0\,\rm{keV})=1.63^{+1.29}_{-0.40}\times 10^{9}\mathrm{\,g\,s^{-1}}$.
This is the only 
measurement of
the instantaneous accretion rate of G29--38 
from X-ray observations
and is the first such measurement at any metal-polluted white dwarf. 

The accretion rate measured from the X-rays is strictly a lower limit on the true accretion rate owing to: (1) limited constraints on the accretion-induced flux outside the observed wavelengths, and (2) the possible contribution of cyclotron cooling of the post-shock plasma.
These two factors 
potentially allow an accretion rate higher than the one measured from the \textit{Chandra} data alone, and we 
consider them 
both
in turn. 

The observations provide robust constraints on hard X-ray emission ($>$2.0\,keV) as the \textit{Chandra} ACIS-S detector is most sensitive in the range 1.0--6.5\,keV (see Extended Data Fig.\,\ref{fg:counts-effective-area}), whereas significant emission is likely softwards of the ACIS-S bandpass ($<0.3{-}0.5$\,keV). 
To
estimate the flux carried at unobserved wavelengths we integrated the best-fit cooling flow model across the wider energy band 0.0136--100\,keV 
(see Methods).
We found the flux in the wider integration band to be $F_{\rm X}(0.0136{-}100\,\mathrm{keV})=4.34^{+14.8}_{-1.72}\times10^{-15}\,\mathrm{erg\,s^{-1}\,cm^{-2}}$, with errors corresponding to the 68\% confidence interval. This represents a flux increase of a factor $2.2^{+7.5}_{-0.9}$ compared with the \textit{Chandra} ACIS-S passband integration, effectively setting the parameter $A=0.5^{+0.3}_{-0.4}$ from Equation\,\eqref{eq:Mdot-Xray}. 

A magnetic field on G29--38 has the potential of funneling the accretion flow towards the magnetic poles of the white dwarf\cite{ghosh1978,metzger12}, and if the density in the post-shock region is sufficiently high, cyclotron emission at radio wavelengths could contribute to its cooling, which would imply that the accretion rate based on the X-ray data is underestimated. 
An estimate for the temperature, $T_{B}$, above which cyclotron emission cooling dominates was given by equation 10 from ref.\cite{farihi2018XRay} such that  
one can approximate the total luminosity to be $L_{\rm tot} \approx (T_{\rm X}/T_B)L_{\rm X}$ for a thermal plasma with temperature $T_{\rm X}$. 
The 3$\sigma$ upper limit on the magnetic field of G29--38 which was found to be 1.5\,kG from spectropolarimetry observations with FORS2\cite{farihi2018XRay}, and even if the magnetic field of G29--38 is close to the observational detection limit, the maximum possible correction due to cyclotron emission cooling is predicted to be a factor $1.1^{+0.6}_{-0.1}$ given the observed X-ray accretion rate (see Extended Data Fig.\,\ref{fg:lum-increase-cyclotron}). And for magnetic fields less than $\approx$1\,kG negligible cyclotron cooling emission is predicted.
In summary, compared to the 
X-ray accretion rate measured in the ACIS-S passband,
$\dot{M_{\rm X}}=1.63^{+1.29}_{-0.40}\times 10^{9}\mathrm{\,g\,s^{-1}}$,
the true accretion rate could be higher by 
a factor 
$1.1^{+0.6}_{-0.1}$ due to possible cyclotron radiation and a factor
$2.2^{+7.5}_{-0.9}$
due to additional flux emitted at unobserved wavelengths (see Methods and Extended Data Table\,\ref{tab:confidence-interval-Flux}).

Accretion rates at metal-polluted white dwarfs are typically inferred indirectly from spectroscopic abundance measurements, coupled with white dwarf atmospheric models that quantify the flux of metals moving through the photosphere\cite{dupuis1992,koester09} in the accretion-diffusion scenario. Using this approach,
the time-averaged accretion rate of G29--38 has previously been inferred\cite{farihi09,xu14} to be $\dot{M}=(5.0 \pm 1.3)\times 10^{8}$ and $(6.5 \pm 1.6) \times 10^{8}\,\mathrm{g\,s^{-1}}$, where the errors indicate the typical ${\approx}25$\% uncertainty on spectroscopic abundance measurements\cite{farihi09}. Fig.\,\ref{fg:Mdot-1D-3D} shows the calculated X-ray accretion rates, and those derived from previous spectroscopic studies.
Our observations establish an X-ray accretion rate which agrees to within a factor 3 of those derived via the spectroscopic method. The key parameters in the determination of a spectroscopic accretion rate are the observed spectroscopic metal abundance, the convectively-mixed mass at the surface, and diffusion timescale at the base of the mixed region (see Equation\,\ref{eq:Mdot-steady-state} in Methods). 
Recent results from 3D radiation-hydrodynamic simulations predict that the convectively-mixed mass can increase by up to 2.5 orders of magnitude when accounting for enhanced mixing due to convective overshoot\cite{freytag96,kupka18,cunningham19}, whilst the diffusion timescale can increase by up to 1.5 dex. 
This leads to a temperature-dependent increase in accretion rate inferred from spectroscopic observations due to convective overshoot\cite{cunningham19}. At the effective temperature of G29--38 ($\approx$11,500--12,000\,K), convective overshoot is predicted to increase the accretion rate by a factor 3--4. Thus the 3D accretion rate is predicted to be $\approx2\times10^9$\,g\,s$^{-1}$, which is in close agreement with the derived X-ray accretion rate (see Fig.\,\ref{fg:Mdot-1D-3D}).
In the atmospheric layers, convective overshoot increases both the convectively-mixed mass and the diffusion timescale of elements settling out of the fully mixed surface layers. At an effective temperature of $\approx$12,000\,K the inclusion of convective overshoot increases the typical predicted diffusion timescale from days to years,  which is more consistent with the lack of time variability of metal lines in G29--38\cite{debes08}.

Our \textit{Chandra} observations of G29--38 establish white dwarfs accreting material from disintegrating planetary bodies as a new class of X-ray source, demonstrating a novel observational technique to investigate evolved planetary systems. In particular, X-ray observations can provide instantaneous measurements of the accretion rate onto the white dwarf, in contrast to the time-averaged estimates derived from model atmosphere analyses, which are, moreover, subject to remaining uncertainties in the assumptions about convection and diffusion. As such, repeated observations of the same system will enable measurements of the temporal variability of the accretion rate from the disk onto the white dwarf, thereby providing direct observational constraints on the evolution of the material within the circumstellar disks\cite{swan19,xu18}, and the physical mechanisms driving the accretion from that disk onto the white dwarf\cite{metzger12,kenyon2017b,farihi2018XRay}. The X-ray fluxes of these systems are low, and only a handful are within the reach of current facilities, however, large-aperture future X-ray missions such as the \textit{Advanced Telescope for High-ENergy Astrophysics}\cite{barret2020Athena} (\textit{ATHENA}) will enable systematic X-ray studies of evolved planetary systems.

\clearpage


\bibliographystyle{naturemag}
\newcounter{firstbib}

\clearpage



\clearpage       

\begin{methods}

\subsection{Observations and data reduction.}

Our observations consisted of five exposures using \textit{Chandra} ACIS-S carried out between 22 and 27 September 2020. Each observation had an exposure time between $\approx$15--25\,ks (see Extended Data Table\,\ref{tab:Chandra-observations}), with a total, merged exposure time of 106.33\,ks, or 29.5\,hr.
All observations were carried out with the ACIS-S instrument using the \textit{Gaia} DR2\cite{gaia2018} proper motion-corrected coordinates of our target positioned on the S3 chip. The predicted X-ray luminosity was sufficiently low that our setup included the very faint mode which stores event grades in regions of $5\times5$ pixels, rather than the standard $3\times3$. This allows an improved identification and rejection of background events, particularly at hard and very soft energies. 

The data reduction was performed using the software package \textit{Chandra} Interactive Analysis of Observations (\texttt{CIAO}\cite{CIAO2006fornaturesubmission}). The reduction began by reprocessing the five observations using the \texttt{chandra\_repro} package with standard grade, status and good time filters and \texttt{vfaint} background cleaning applied. We employed \texttt{merge\_obs} to merge the five observations into a single set of event files along with images, point spread function (PSF) maps and exposure maps. We performed this routine using combinations of the standard \textit{Chandra} science energy bands: soft (0.5--1.2 keV), soft+medium (0.5--2.0 keV) and broad (0.5--7.0 keV). 
During each merging routine, the PSF maps and exposure maps were generated for each band with exposure evaluated on 0.9 keV, which represents an approximate mean of the suspected source photon energies. 

\subsection{Statistical significance of detection}

To determine the statistical significance of the detection, 
we utilised the Poisson distribution, as appropriate for counting statistics with small numbers of events\cite{LiMa1983poisson}. This allows us to quantify the 
confidence with which we can rule out the null hypothesis that we detected no source photons. We defined a source region of 1\,arcsec around the coordinates of our target -- Right Ascension and Declination of
$(352.196185 \pm 0.00004)$\textdegree\ and $(+05.24686 \pm 0.00003)$\textdegree,
respectively -- which have been corrected to the J2020.73 epoch using the \textit{Gaia} EDR3\cite{gaia2021EDR3} astrometry. The nearby background region was defined as a circle with a 52\,arcsec radius, avoiding obvious point sources -- such as the nearby bright source which contaminated previous \textit{XMM-Newton} observations\cite{jura09} (e.g., see Extended Data Fig.\,\ref{fg:wavdetect-output-6-panel}; panels a--c) -- as well as drops in sensitivity in the exposure map.

Extended Data Table\,\ref{tab:Poisson-source-and-background} shows the source and background counts from the full, merged observation of 106.33\,ks. The Poisson distribution gives the probability of receiving $N$ or more counts, given an expected mean background, $b$, as
\begin{equation}
	P(N|b) = \frac{b^N e^{-b}}{N!} \, .
\end{equation}
We can therefore reject the null hypothesis that there are no source photons with a confidence level ($\mathrm{CL}$) of $\mathrm{CL}=100\times(1-P(N|b))$. In the soft, soft+medium and broad bands, we find $\mathrm{CL}=$\,99.99962\%, 99.99975\%, and 99.97255\%, respectively. The statistical significance, $s$, can be computed as
\begin{equation}
 s=\mathrm{erf}^{-1}\left(1-P(N|b)\right)\sqrt{2}, 
\end{equation}
which for each of the three bands is $s=$\,4.62, 4.71, and 3.64$\sigma$, respectively. 
We note that adopting a smaller aperture with a radius of 0.5\,arcsec, which is appropriate for soft sources with ACIS-S (see figure 6.10 of the \chandra\ Proposers' Observatory Guide Version 23.0\footnote{https://cxc.harvard.edu/proposer/POG/pdf/MPOG.pdf} where the enclosed fractional power is $>$50\% for 0.5 arcsec apertures), and accounting for the astrometric uncertainty by centring the aperture on the locus of photons at the source position (see Fig.\,\ref{fg:RA-DEC-events-uncertainty}), increases the detection significance in the three science bands to 5.65, 5.94 and 5.06$\sigma$, respectively.
We conclude that in all three science bands considered, 
an X-ray source is detected at the expected position of our target, G29-38.

\subsection{Count rates and confidence intervals}
For each science band considered, we converted the number of counts to a count rate using the total exposure time of 106.33\,ks. In the soft, soft+medium and broad bands we find count rates of $0.038^{+0.017}_{-0.022}$, $0.047^{+0.020}_{-0.023}$ and $0.047^{+0.024}_{-0.019}$\,ks$^{-1}$ per 1\,arcsec aperture, respectively. The 68\% confidence interval on count rate across all three bands is $(2.1$--$7.0)\times10^{-5}$ counts per second, while the 90\% confidence interval spans $(1.3$--$9.0)\times10^{-5}$ counts per second. The confidence intervals were calculated using a Bayesian approach to Poisson statistics, following methodology in ref.\cite{kraft1991}. This uses a simple prior which requires a non-negative number of source counts.

\subsection{Investigating positional uncertainty}
The five recorded soft+medium band events within 1\,arcsec of our target coordinates have a sky position which is
shifted by approximately $-0.5$ arcsec in right ascension. In Fig.\,\ref{fg:RA-DEC-events-uncertainty} we plot the sky position of the recorded events in the soft+medium band within 1.1\,arcsec of our target. We find that all five of the photons in the soft+medium bands lie within 1$\sigma$ of the target's position on the CCD. This confidence interval comprises two sources of uncertainty. 
The first is from the four \gaia\ astrometric parameters, i.e., coordinates (R.A. and DEC) and proper motion ($\mu_{\rm ra}$ and $\mu_{\rm dec}$). The positional error at the epoch of our \chandra\ observations, $x$, is thus propagated as 
\begin{equation}
\delta x_{i} =[(t \delta \mu_{i})^2 + (\delta x_{0,i})^2]^{1/2}\, ,
\end{equation}
where $t$ is the elapsed time between the \gaia\ epoch and the \chandra\ observations, $\mu$ is the proper motion, $x_0$ is the \gaia-measured coordinate, and the subscript $i$ indicates either R.A. or DEC.

The second positional uncertainty arises from the astrometric uncertainty of the telescope itself. From figure 5.4 of the \chandra\ Proposers' Observatory Guide (Version 23.0)\footnote{https://cxc.harvard.edu/proposer/POG/pdf/MPOG.pdf}, the 68\% confidence limit on the radial offset of a given target is 0.5\,arcsec. This value was computed for observations between 2015--2020, by comparing the radial offset of \chandra\ sources with optical sources predominantly from the Tycho-2 catalog. 
Combining these two sources of positional uncertainty, 
the thick ellipse in Fig.\,\ref{fg:RA-DEC-events-uncertainty} shows the 68\% confidence interval on the position of our target
and illustrates the expected precision of the target position in our data.
We find that all five events considered in our earlier statistics fall within this area.

\subsection{Ruling out background contamination}
To determine the probability of chance alignment with a background source we follow two methodologies. In the first,
we run a source-detection algorithm to provide an independent constraint on the number of sources at this depth of pointing.
We rely upon \texttt{wavdetect}\cite{freeman2002wavdetect}, a wavelet-based algorithm included in the \texttt{CIAO} package, designed for the spatial analysis of Poisson count data. As recommended in the documentation, we set the significance threshold parameter, \texttt{sigthresh}, to $1/n_{\rm px}\approx 10^{-6}$, which implies that approximately one identified source will be a false detection. We perform this analysis in all three of the standard bands considered in this study, as well was for two values of \texttt{sigthresh}, $s_1$ and $s_2$. In the first, we use $s_1=10^{-6}$, which is the recommended value for the number of pixels (1024$\times$1024) of the S3 CCD. The second, more stringent, value of \texttt{sigthresh}, $s_2=5\times 10^{-7}$, accounts for the `dead' corners of the image which arise from aligning the CCD image with the world coordinate system (WCS). This results in an image with 1414$\times$1401 pixels. The results are given  in Extended Data Table\,\ref{tab:wavdetect-results-table}, and Extended Data Fig.\,\ref{fg:wavdetect-output-6-panel} shows the sky location of the detected sources in all three science bands. Across the entire CCD image we detect 7, 25 and 32 sources, in the soft, soft+medium and broad bands, respectively. We find our target is not detected in the broad band image, regardless of the \texttt{sigthresh} value. In the soft band, the target is detected at the higher \texttt{sigthresh} value, but not with the more stringent value. And in the soft+medium band we find the source is detected in both cases. Dividing the total number of sources in each band by the field of view ($1024^2 \times 0.4920^2$\,arcsec$^2$) reveals sky densities of $2.76$, $9.85$ and $12.6\times 10^{-5}$\,arcsec$^{-2}$, in the soft, soft+medium and broad bands, respectively. Interpreting this as the probability of chance alignment with a background source, 
we can rule out chance alignment at a significance of $\mathrm{erf}^{-1}(1-\rho_{\rm sky})\sqrt{2} = $\,4.2, 3.9 and 3.8$\sigma$, or confidence of 99.997, 99.990 and 99.987\%, respectively.

To further test this result, we performed a Monte Carlo aperture photometry experiment.
We placed 100,000 apertures with 1\,arcsec radii within 100\,arcsec of the target. This area was chosen to confine the study to the region of the ACIS-S3 CCD where the PSF does not exhibit a significant gradient, as beyond $\approx$120\,arcsec the PSF increases significantly. This effect can be seen in the output of the \texttt{wavdetect} algorithm in the top panels of Extended Data Fig.\,\ref{fg:wavdetect-output-6-panel}, where the sources detected near the edge of the CCD appear to be much larger. For the Monte Carlo aperture photometry test we make use of the \texttt{Python} package \texttt{scipy}\cite{scipy2020}, in particular the \texttt{KDTree} module. This facilitates the rapid comparison of a large number of test aperture positions with those of recorded events from our observation. To estimate the true sky density of background sources we exclude the target and known nearby bright source using circular masks of radius 1\,arcsec and 2.5\,arcsec, respectively. The final number of test apertures was slightly reduced after the removal of the $\approx$150--200 which fell within masked regions. 

The results of this test on the soft and soft+medium band events are shown in Extended Data Fig.\,\ref{fg:ap-throw-down-soft-medium-mask}.
In the soft band analysis, we find 1 in 99,828 test apertures (0.001\%) retrieved four counts, while no test apertures found more than four counts.
This allows us to rule out chance alignment with a background source at a significance of 4.4$\sigma$.
In the soft+medium band, we find 56 in 99,852 test apertures (0.056\%)
have five or more counts, while 48 (0.048\%) returned between 6--27 counts, allowing us to rule out chance alignment at the 3.4$\sigma$ level. 
The reason the significance is lower in the soft+medium band is not because there is any doubt about the source detection, but because (1) our source has a soft spectrum and so contributes little flux in the medium band, and (2) there are a much larger number of background sources in the medium band, since most X-ray sources are harder than our target, and because the effective area of the telescope is higher.
This test also provides an empirical confirmation of the quoted expected backgrounds in Extended Data Table\,\ref{tab:Poisson-source-and-background}, with fewer than 10\% and 20\% of 1\,arcsec test apertures returning any counts in the soft and soft+medium bands, respectively.

We have shown that the counts detected within 1\,arcsec of our target coordinates allow the rejection of the null hypothesis  -- that no source counts were measured -- with a conservative confidence of 99.97255--99.99962\% (3.64--4.71$\sigma$). 
Using the recommended aperture size of 0.5\,arcsec, and centring the aperture on the observed source position, we find the significance increases to 5.06--5.91$\sigma$. We have also shown that the observed position of the five recorded events is consistent with the
expected position of our target, to within the 1$\sigma$ astrometric uncertainty. We have tested the hypothesis that recorded source counts could have originated instead from a background source, and by constraining the sky density of sources at this depth of pointing using \texttt{wavdetect}, we rule out a chance alignment with a background source 
at a confidence of 99.997 and 99.990\% (4.2 and 3.9$\sigma$) in the soft and soft+medium bands, respectively. 
Confirming these results with Monte Carlo aperture photometry, in the soft band, with 0.001\% of test apertures returning four counts, chance alignment can be ruled out with a confidence of 99.999\% (4.4$\sigma$). 
Given the high degree of confidence to which the observed events can be attributed to our target, G29--38, in the following we perform spectral modelling to derive an X-ray flux, luminosity and accretion rate.

\subsection{Spectral model and debris composition}
We perform a spectral analysis assuming two different optically-thin plasma models, and three distinct abundance profiles. The first model is a one-component, isothermal plasma
implemented in the \texttt{vvapec} model within the \texttt{XSPEC} software package (version 12.11.1),\cite{arnaud1996XSPECfornaturesubmission} which uses the AtomDB atomic database\cite{foster12}. The second model, \texttt{mkcflow}, is a cooling flow model that allows for a 
range
of temperatures, with the relative emission measure for each temperature weighted by the inverse of its emissivity. We selected the option to use the AtomDB database. This cooling flow model is more physically-motivated than the isothermal model and is likely to provide a better estimate of the X-ray flux beyond the observed bandpass. 
We fitted the \texttt{mkcflow} model
by fixing the lowest temperature plasma at the lower limit of the model of $kT=0.08$\,keV, 
only allowing the 
upper temperature of the range 
to vary, enabling a one-parameter fit. This models the emission from material heated to the upper temperature and then cooling to temperatures lower than can be detected with \chandra\ ACIS-S. The three abundance profiles we use are Solar\cite{asplund2009}, bulk Earth\cite{mcdonough1995}, and the observed photospheric metal abundances\cite{xu14} of G29--38 with an equal number abundance of hydrogen. The composition of the infalling material is best described as a rocky, water-depleted, chondritic object\cite{xu14}, with detected lithophile (O, Si, Mg, Ca, Ti, and Cr), siderophile (Fe), and atmophile (C) elements. If the 
heated
plasma is formed sufficiently close to the stellar surface, such as in the bombardment scenario, the rocky accreted material may be mixed with photospheric hydrogen. However, we found that our results were not sensitive to the hydrogen abundance, as at the best-fit plasma temperatures ($\approx$0.5\,keV) the cooling is dominated by metal line emission. At the distance to G29--38, the interstellar column density is expected\cite{redfield00} to be only around $N_{\rm H}=5.4\times10^{18}\rm\,cm^{-2}$, which has a negligible effect in the \chandra\ ACIS-S passband (${<}0.5$\% absorption at 0.5\,keV). The five events in the ACIS-S spectrum were fitted unbinned, using the C-statistic,\cite{cash79} and with no background subtraction. 

Extended Data Figure \ref{fg:xspec-fits} shows the spectral fits and Extended Data Table\,\ref{tab:confidence-interval-kT} shows the best-fit plasma temperatures for all six models, as well as the 68\% and 90\% confidence intervals. 
The bulk Earth and photospheric models  agree to within  the 68\% confidence interval. The observed photospheric abundances of G29--38 could be scaled by the microscopic diffusion timescale in the atmosphere to infer a more accurate accreted debris composition. The difference, however, would be small, and the agreement between the bulk Earth and photospheric models strongly suggests that the observations would not be sensitive to such a correction. 

\subsection{Deriving X-ray Flux}
We derive a total flux due to accretion by integrating the best-fit spectral models over a finite frequency (or energy) range. We find a robust lower limit on the X-ray flux of 
$F_{\rm X}(0.3{-}7.0\,\rm{keV})=1.97^{+1.55}_{-0.48}\times 10^{-15}\,\mathrm{erg\,s^{-1}\,cm^{-2}}$ 
by performing the integration only over the energies within the \textit{Chandra} ACIS-S passband (0.3--7.0\,keV). We also perform the integration across a slightly narrower band (0.5--7.0\,keV) as the instrument sensitivity below 0.5\,keV has degraded since launch and is now relatively low. 
These results, with their associated confidence intervals, are shown in Extended Data Table\,\ref{tab:confidence-interval-Flux}.

With the lower limit on X-ray flux tightly constrained, providing an upper limit on the X-ray flux is more challenging. This is primarily due to the lack of observations between the very soft X-rays ($\sim$0.1\,keV) and the UV. There are no instruments currently equipped to perform observations at these Extreme Ultraviolet (EUV) wavelengths, so directly measuring the flux in this regime is, for now, impossible. 
The constant $A$ included in Equation\,\eqref{eq:Mdot-Xray} describes the predicted fraction of the total luminosity carried in the \textit{Chandra} ACIS-S passband.
Previous studies\cite{jura09,patterson85} have made the approximation that $A=0.5$, or $0.25$. 
We provide an upper estimate of the X-ray and EUV flux by integrating the best-fit spectral models over a much wider energy range (0.0136--100.0\,keV). Model spectra over this broadband are plotted in Extended Data Fig.\,\ref{fg:xuv-spectrum} and we include fluxes calculated for this broad band in 
Extended Data Table\,\ref{tab:confidence-interval-Flux}. 

Extended Data Fig.\,\ref{fg:Mdot-1D-3D-mkcflow-3panel} shows the X-ray flux computed in the three integration bands used in this analysis. For clarity, from here onwards we show results only from the cooling flow model (\texttt{mkcflow}) using the photospheric abundances. The use of the cooling flow model is designed to provide a realistic temperature distribution and hence realistic fluxes in the EUV band (it can be compared with the isothermal model in Extended Data Fig.\,\ref{fg:xuv-spectrum}). From left to right in Extended Data Fig.\,\ref{fg:Mdot-1D-3D-mkcflow-3panel} the size of the integration domain increases, and as expected the best-fit X-ray flux (and thus accretion rate) also increases, with the widest band (0.0136--100\,keV) providing an upper estimate of the X-ray and EUV flux. The wider integration band results in a best-fit accretion rate which is $2.2^{+7.5}_{-0.9}$ times higher than that derived from the ACIS-S passband, where the errors represent the 68\% confidence interval. We note that the 68\% and 90\% upper bounds on the fluxes increase by a factor $\sim$10 and $\sim$100, respectively, due to the uncertain importance of the EUV band.  

\subsection{X-ray luminosity} We derive an X-ray luminosity from the X-ray flux measured in the previous section using $L_{\rm X}=4\pi d^2 F_{\rm X}$ with the distance to G29--38 of
$d=(17.53\pm0.01)$\,pc
calculated from the  \textit{Gaia} EDR3 parallax. Extended Data Table\,\ref{tab:confidence-interval-luminosity} shows the best-fit X-ray luminosity for each of the energy ranges. From the robust lower limit on the X-ray flux, the best-fit 
X-ray luminosity from our observations is found to be 
$L_{\rm X}(0.3-7.0\,{\rm keV)}=7.24^{+5.66}_{-1.76}\times10^{25}\,\mathrm{erg\,s^{-1}}$,
where the errors are indicative of the 68\% confidence interval.

\subsection{X-ray accretion rates}
Here we estimate the instantaneous accretion rate using a simple model which has been employed in studies of accreting white dwarfs\cite{patterson85,jura09,farihi2018XRay}. In this model, infalling material reaches near-free fall velocity, forming a heated plasma as the material approaches the white dwarf surface\cite{mukai2017}. The plasma must radiate outwards a total energy equivalent to the initial gravitational potential energy of the accreted material such that\cite{patterson85}
\begin{equation}
	L_{\rm tot} = \frac{1}{2}\frac{GM_{\rm WD}\dot{M}}{R_{\rm WD}},\,
\end{equation}
where $R_{\rm WD}$ is the white dwarf radius, at which infalling material impacts the atmosphere,
and the factor 1/2 accounts for 50\% of emitted photons being directed back towards and absorbed by the star\cite{patterson85,kylafis1982}.
We allow the emitting plasma to be formed by individual atoms reaching the white dwarf atmosphere in a scenario termed the ``bombardment'' solution, relevant for low accretion rate systems,\cite{kuijpers1982,woelk1993}
such that the X-ray emitting plasma is formed at the white dwarf radius ($R_{\rm WD}=0.0129R_{\odot}$).
This scenario was hypothesised for white dwarfs accreting from main-sequence companions at low accretion rates, and predicts a plasma temperature ($\approx0.6$\,keV) comparable to that which we have measured from our observations ($0.5$\,keV).
From the total luminosity, the X-ray luminosity can be written $L_{\rm X}=AL_{\rm tot}$, where the constant $A$ accounts for the fraction of the flux emitted outside the observed passband. 
If the plasma cooling is entirely mediated by line cooling in the observed X-ray passband (0.3--7.0\,keV), then $A=1$. If the plasma experiences additional line cooling at harder, or softer energies than those observed, then $A<1$. If the plasma cools via other physical mechanisms, such as cyclotron emission cooling, then some of the total luminosity will be radiated at radio wavelengths, in which case $A<1$. This has typically been set to $A=0.25{-}0.5$ in previous X-ray studies\cite{patterson85,jura09}. The flux integration over the wider energy band (0.0136--100\,keV) allowed us to constrain the increase due to unobserved flux to a factor $2.2^{+7.5}_{-0.9}$, which corresponds to $A=0.5^{+0.3}_{-0.4}$, consistent with estimates from previous studies for white dwarfs accreting from main-sequence companions, with errors corresponding to the 68\% confidence interval.

The accretion rate can be inferred from the \textit{Chandra} ACIS-S observations using Equation\,\eqref{eq:Mdot-Xray}, which describes an accretion flow converting its gravitational potential energy into an X-ray luminosity.
In Extended Data Table\,\ref{tab:confidence-interval-Mdot} we use this equation in the limiting case, with all X-ray flux emitted within the \textit{Chandra} ACIS-S passband ($A=1$)
to transform the X-ray luminosities into accretion rates. 
We find the X-ray accretion rate, measured from our observations, to be
$\dot{M_{\rm X}}(0.3{-}7.0\,\rm{keV})=1.63^{+1.29}_{-0.40}\times 10^{9}\mathrm{\,g\,s^{-1}}$.
The true accretion rate could be inferred to be higher for two reasons:
firstly, there may be additional flux carried at lower, unobserved energies, and secondly, there could be some contribution if the magnetic field is sufficiently high to promote some cooling via cyclotron emission cooling\cite{farihi2018XRay}.
We have constrained the first consideration by integrating the best-fit model spectra over the full EUV and X-ray regime, finding the flux to increase by up to a factor $2.2^{+7.5}_{-0.9}$, or $A=0.5^{+0.3}_{-0.4}$. We have robust constraints on the second consideration, finding a maximum increase due to cyclotron emission cooling of a factor
$1.1^{+0.6}_{-0.1}$ (see Extended Data Fig.\,\ref{fg:lum-increase-cyclotron}).
Combining both possibilities for an increase in accretion rate, the upper estimate on the true accretion rate could be a factor 
$2^{+14}_{-1}$
higher than the observed X-ray accretion rate, where the errors represent the 68\% confidence interval.

\subsection{Atmospheric parameters}
G29--38 was identified as a metal-polluted, hydrogen atmosphere (DAZ) white dwarf\cite{koester97}. It is also a pulsating, or ZZ Ceti, star with large-amplitude non-radial pulsations with periods on the time scale $t\approx 110-1250$\,s\cite{patterson1991,kleinmann1998}.
From fitting the broadband energy distribution\cite{mccleery2020,gentile2021}, the effective temperature of G29--38 has been estimated to be $T_{\rm eff}= (11529\pm206)$\,K and $(12090\pm229)$\,K, when using \textit{Gaia} EDR3 and PanSTARRS1 (PS1) photometry, respectively. From the \textit{Gaia} EDR3 parallax, the mass was found to be $M=(0.629\pm0.016)\,M_{\odot}$ and $(0.670\pm0.017)\,M_{\odot}$, when using \textit{Gaia} EDR3 and PS1 photometry, respectively.
This spans the full range of effective temperatures previously derived for this star using the spectroscopic method to fit the Balmer absorption lines\cite{koester97,farihi09,gianninas11,xu14}. Using the observed Balmer line spectrum time-averaged over multiple pulsation cycles\cite{gianninas11} and the latest model spectra\cite{tremblay09} with 3D corrections\cite{tremblay13c} we obtain $T_{\rm eff}= (11906\pm190)$\,K and $M=(0.689\pm0.031)\,M_{\odot}$. 
The model accretion rates (solid lines) derived in Fig.\,\ref{fg:Mdot-1D-3D} are plotted over an x-axis range to include both the EDR3 and PS1 effective temperatures, for a H-atmosphere white dwarf with a mass of $M=0.60\,M_{\odot}$. Radiation-hydrodynamic simulations of convective overshoot are currently only available for H-atmosphere white dwarfs with surface gravities of $\log g=8.0$ ($M_{\rm WD}=0.6M_{\odot}$). For self-consistency, the calculation of X-ray accretion rates (Equation\,\ref{eq:Mdot-Xray}) and spectroscopic accretion rates (Equation\,\ref{eq:Mdot-steady-state}) use the canonical white dwarf mass of $M_{\rm WD}=0.6M_{\odot}$. This is 3--15\% smaller than the mass of G29--38 derived from photometric and spectroscopic fits ($M_{\rm WD}=0.629{-}0.670M_{\odot}$). Given that the X-ray accretion rate scales proportionally with white dwarf mass, this represents only a small uncertainty in the measured accretion rate.

\subsection{Spectroscopic accretion rates}
In the accretion-diffusion scenario, under the steady-state assumption, an accretion rate can be inferred from spectroscopic observations as\cite{dupuis1992,koester09}
\begin{equation}
    \dot{M_i}=X_i \frac{M_{\rm cvz}}{t_{{\rm diff}, i}}\, ,
    \label{eq:Mdot-steady-state}
\end{equation}
where $X_i$ is the photospheric abundance of element $i$, $M_{\rm cvz}$ is the convectively-mixed mass and $t_{{\rm diff}, i}$ is the diffusion timescale set by the microscopic physics at the base of the mixed surface layers.
From spectroscopic observations and model atmosphere analyses, the time-averaged accretion rate of G29--38 has previously been inferred\cite{farihi09,xu14} to be $\dot{M}=(5.0 \pm 1.3)\times 10^{8}$ and $(6.5 \pm 1.6) \times 10^{8}\,\mathrm{g\,s^{-1}}$. The uncertainty on the first value\cite{farihi09} was given as a typical 25\% uncertainty (0.1\,dex) on the spectroscopically measured calcium abundance. The error on the second value\cite{xu14} was also estimated in the same way. 
Our observations establish an X-ray accretion rate which agrees to within a factor 3 of previously derived spectroscopic accretion rates.
Recent results from 3D radiation-hydrodynamic simulations predict a temperature-dependent increase in accretion rate inferred from spectroscopic observations due to convective overshoot\cite{cunningham19}. At the effective temperature of G29--38 ($\approx$11,500--12,000\,K), the accretion rate correction is predicted to increase by a factor 3--4. Thus the 3D accretion rate is be predicted to be $\approx2\times10^9$ g\,s$^{-1}$, which is in
agreement with the derived X-ray accretion rate (see Fig.\,\ref{fg:Mdot-1D-3D}).

\newpage

\end{methods}

\clearpage
\begin{addendum}
\item[Data Availability] The data that support the plots within this paper and
other findings of this study are available from the \textit{Chandra Data Archive}. The observation ID numbers are given in Extended Data Table\,\ref{tab:Chandra-observations}.

\item[Code Availability]The official \textit{Chandra} reduction software package \texttt{CIAO} -- which includes \texttt{merge\_obs}, \texttt{wavdetect} and \texttt{XPSEC} -- is freely and publicly available (\url{cxc.cfa.harvard.edu/ciao/}). So too is the \texttt{Python} package \texttt{scipy}.

\end{addendum}

\clearpage


\clearpage


\begin{addendum}
\item[Author Correspondence] All correspondence regarding this work should be
directed to T. Cunningham (email: timothy.cunningham@warwick.ac.uk).
\item
This research was supported by a Leverhulme Trust Grant (ID RPG-2020-366). PJW, BTG and PET were supported by the UK STFC consolidated grant ST/T000406/1. PET received funding from the European Research Council under the European Union’s Horizon 2020 research and innovation programme number 677706 (WD3D). BTG was supported by a Leverhulme Research Fellowship, OT was supported by a Leverhulme Trust Research Project Grant and FONDECYT project 32103, and DV gratefully acknowledges the support of the STFC via an Ernest Rutherford Fellowship (grant ST/P003850/1).

This research has made use of data obtained from the \textit{Chandra Data Archive} and the \textit{Chandra Source Catalog}, and software provided by the Chandra X-ray Center (CXC) in the application packages \texttt{CIAO} and \texttt{Sherpa}.

This work has made use of data from the European Space Agency (ESA) mission
{\it Gaia} (\url{https://www.cosmos.esa.int/gaia}), processed by the {\it Gaia}
Data Processing and Analysis Consortium (DPAC,
\url{https://www.cosmos.esa.int/web/gaia/dpac/consortium}). Funding for the DPAC
has been provided by national institutions, in particular the institutions
participating in the {\it Gaia} Multilateral Agreement.

The Pan-STARRS1 Surveys (PS1) and the PS1 public science archive have been made possible through contributions by the Institute for Astronomy, the University of Hawaii, the Pan-STARRS Project Office, the Max-Planck Society and its participating institutes, the Max Planck Institute for Astronomy, Heidelberg and the Max Planck Institute for Extraterrestrial Physics, Garching, The Johns Hopkins University, Durham University, the University of Edinburgh, the Queen's University Belfast, the Harvard-Smithsonian Center for Astrophysics, the Las Cumbres Observatory Global Telescope Network Incorporated, the National Central University of Taiwan, the Space Telescope Science Institute, the National Aeronautics and Space Administration under Grant No. NNX08AR22G issued through the Planetary Science Division of the NASA Science Mission Directorate, the National Science Foundation Grant No. AST-1238877, the University of Maryland, Eotvos Lorand University (ELTE), the Los Alamos National Laboratory, and the Gordon and Betty Moore Foundation.

\item[Author contributions] 
T.C. performed most of the data analysis and led the writing of the manuscript. P.J.W. contributed to the original observing proposal, the data analysis and the writing of the manuscript. P-E.T and B.T.G. contributed to the writing of the manuscript and discussions on the various constraints on accretion rates. G.W.K., O.T. and D.V. contributed to the original observing proposal and to discussions throughout the project.

\item[Author Information] Reprints and permissions information is available at
www.nature.com/reprints. Correspondence and requests for materials should be
addressed to T.C. (email: timothy.cunningham@warwick.ac.uk).

\item[Competing Interests] The authors declare that they have no competing
interests.

\end{addendum}
\clearpage


\begin{efigure}
    \RaggedRight
    \begin{tabularx}{\columnwidth}{XXX}
        \textsf{\textbf{a}} & \textsf{\textbf{b}} & \textsf{\textbf{c}}
    \end{tabularx}
  \centering
  \justifying
    \vspace{-14pt}
    \includegraphics[angle=0,width=.325\textwidth]{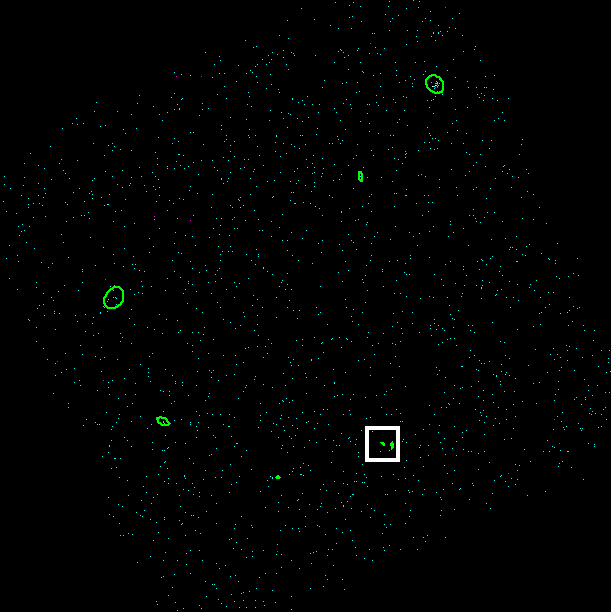}	\hfill
    \includegraphics[angle=0,width=.325\textwidth]{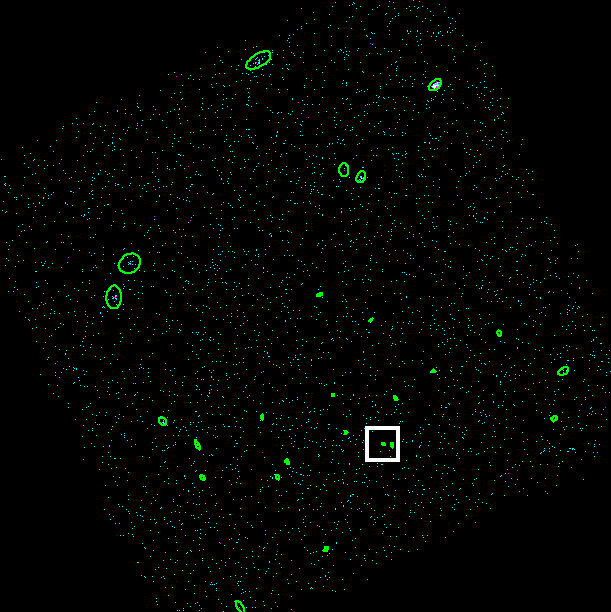} \hfill
    \includegraphics[angle=0,width=.325\textwidth]{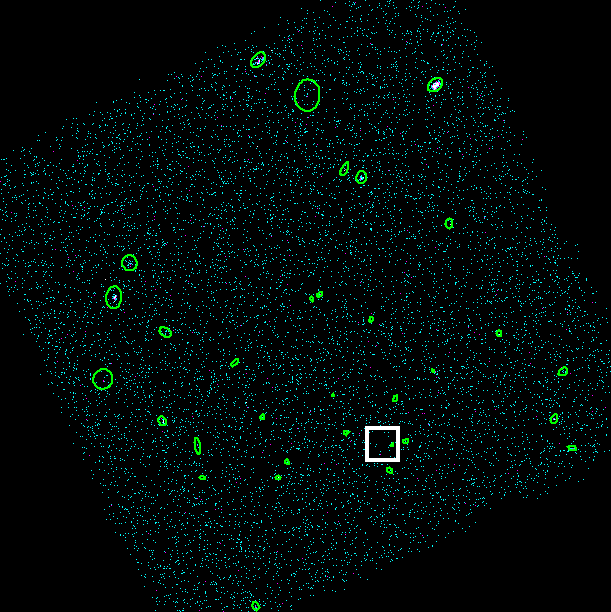} 
    \RaggedRight
    \begin{tabularx}{\columnwidth}{XXX}
        \textsf{\textbf{d}} & \textsf{\textbf{e}} & \textsf{\textbf{f}}
    \end{tabularx}    
    \justifying
    \includegraphics[angle=0,width=.325\textwidth]{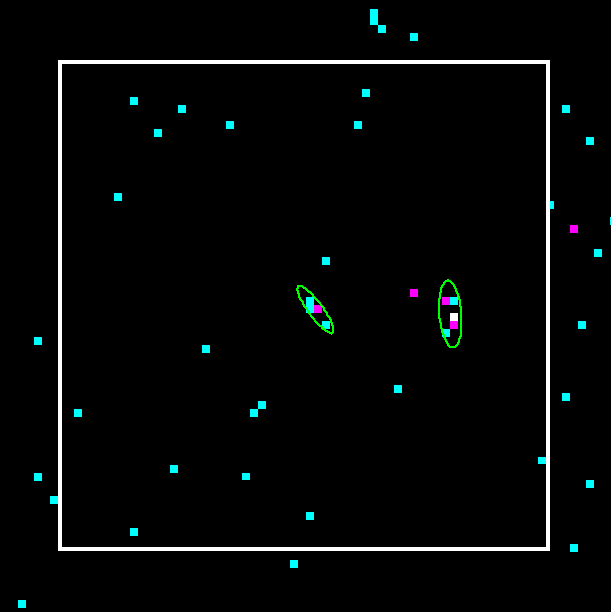} \hfill
    \includegraphics[angle=0,width=.325\textwidth]{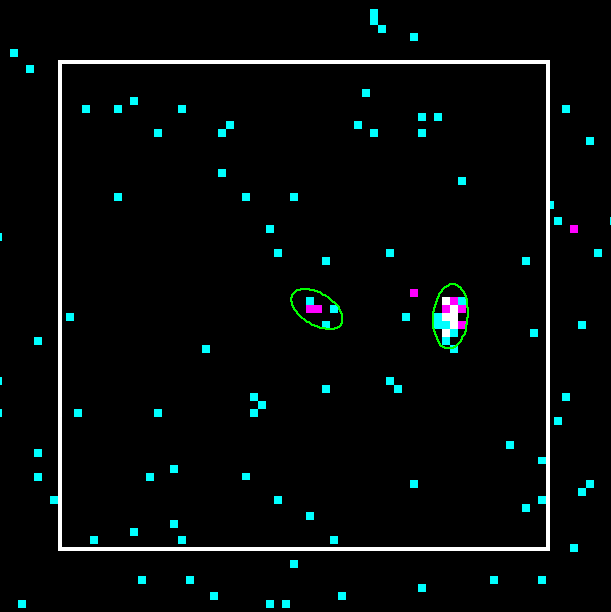} \hfill
    \includegraphics[angle=0,width=.325\textwidth]{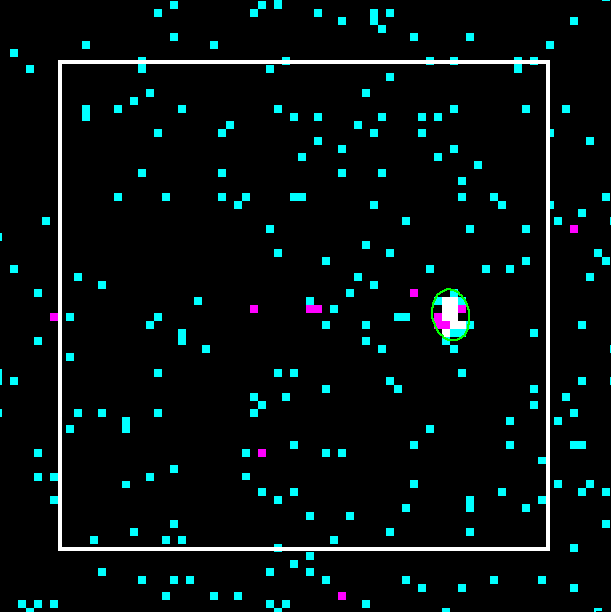} \\
  \caption{\textbf{Estimate of sky density from a source detection algorithm.} Output of the \texttt{wavdetect} source detection algorithm using the recommended significance threshold, \texttt{sigthresh}, of $s_1=10^{-6}$. From left-to-right are the results for the science bands used in this study; soft, soft+medium and broad, with 1, 2, and 3 counts shown in cyan, magenta, and white, respectively. The top panels show the full field of view of the S3 chip on ACIS-S with sources identified by \texttt{wavdetect} shown in green. The bottom panels give a magnified view of the vicinity near the target, where the white square has sides of length 30 arcsec and is centered on the target coordinates. The source at the sky location of G29-38 is detected in the soft and soft+medium band images. The number of sources and corresponding sky density for each band can be found in Extended Data Table\,\ref{tab:wavdetect-results-table}.}
  \label{fg:wavdetect-output-6-panel}
\end{efigure}

\begin{efigure}
    \RaggedRight
    \begin{tabularx}{\columnwidth}{XXXXX}
        \textsf{\textbf{a}} & & \textsf{\textbf{b}} & &
    \end{tabularx}
    \justifying
  \centering
      \vspace{-20pt}
    \includegraphics[angle=0,width=.4\textwidth]{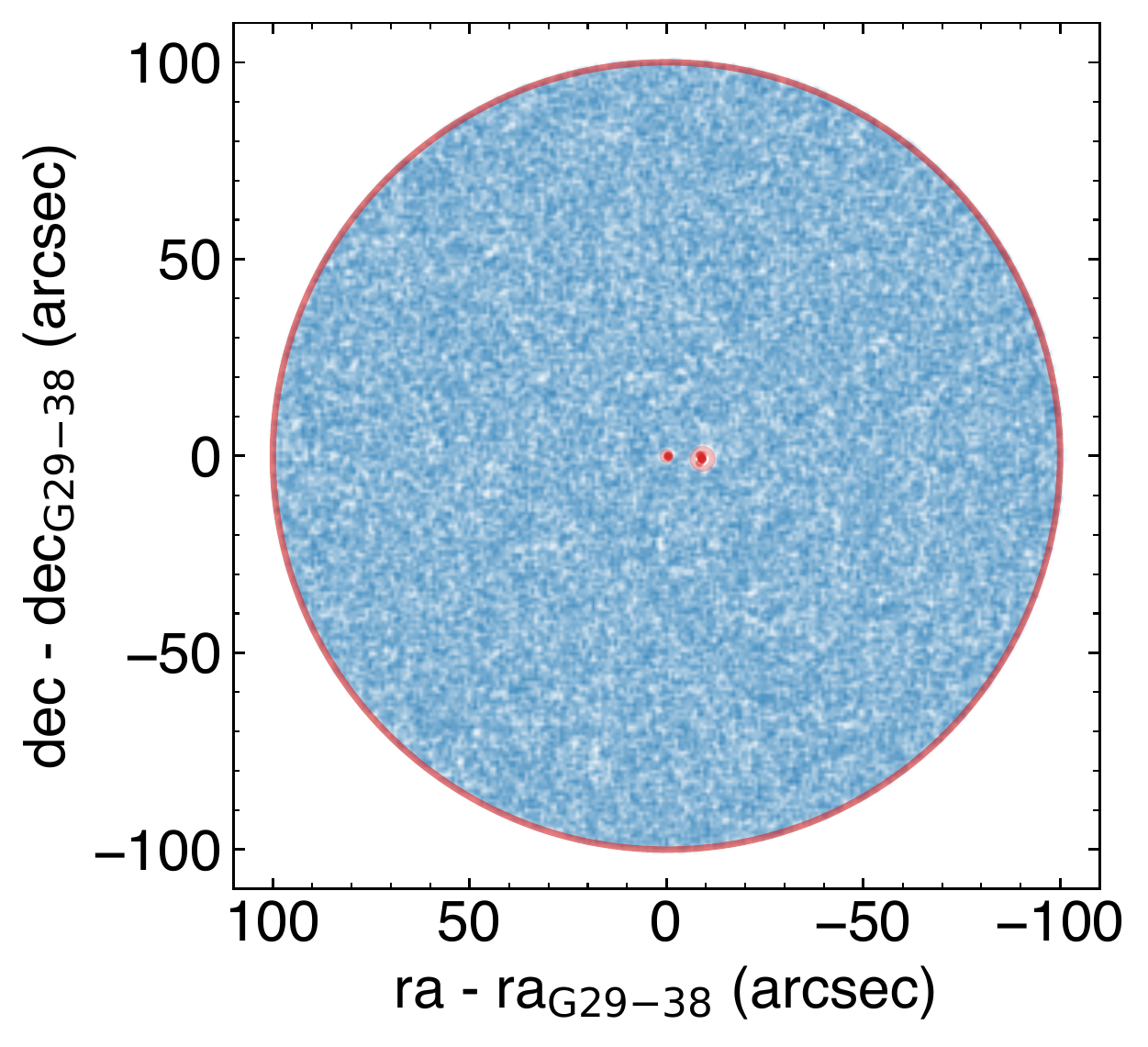}
    \includegraphics[angle=0,width=.59\textwidth]{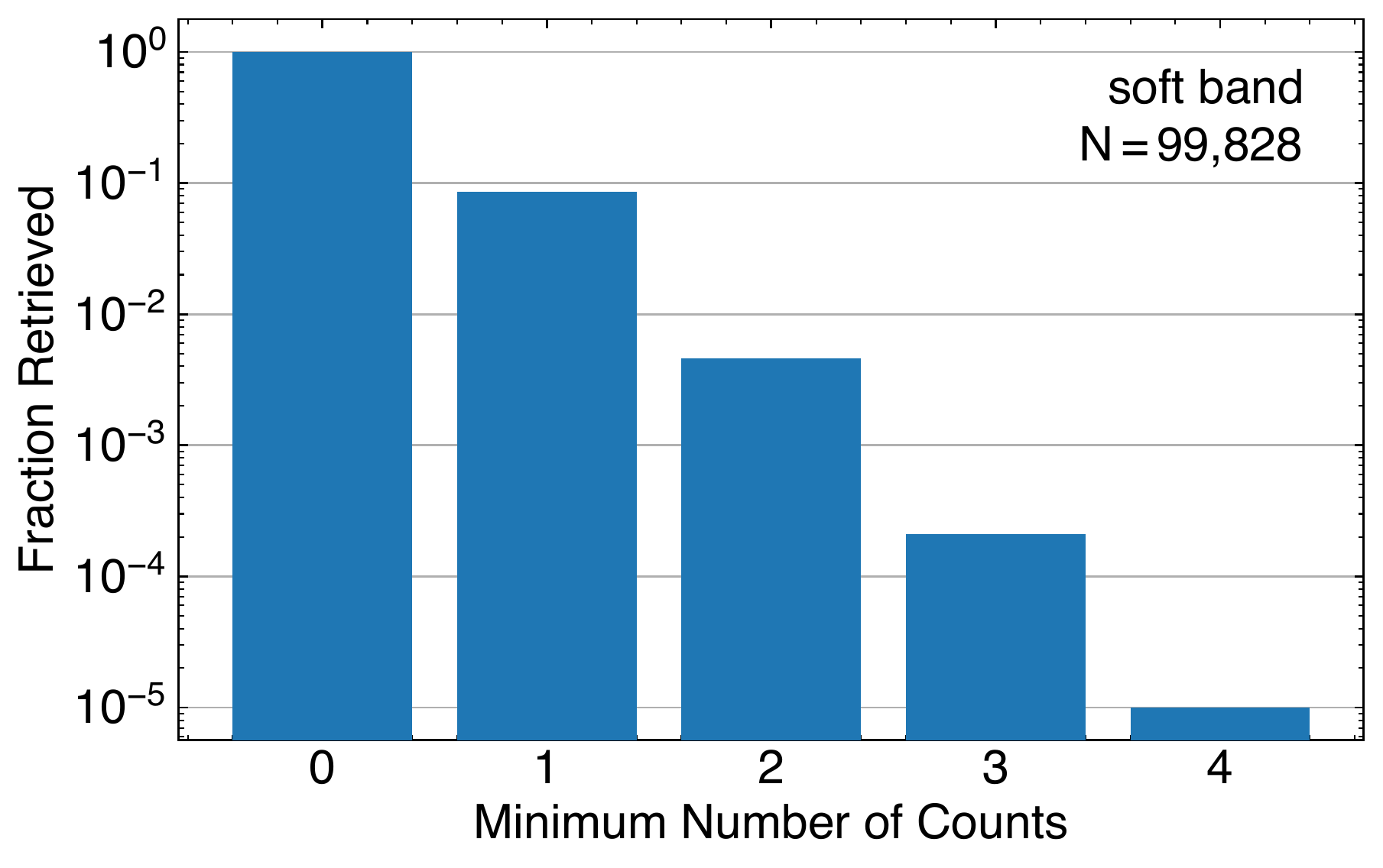}
  \RaggedRight
    \begin{tabularx}{\columnwidth}{XXXXX}
        \textsf{\textbf{c}} & & \textsf{\textbf{d}} & &
    \end{tabularx}
    \justifying
  \centering
    \includegraphics[angle=0,width=.4\textwidth]{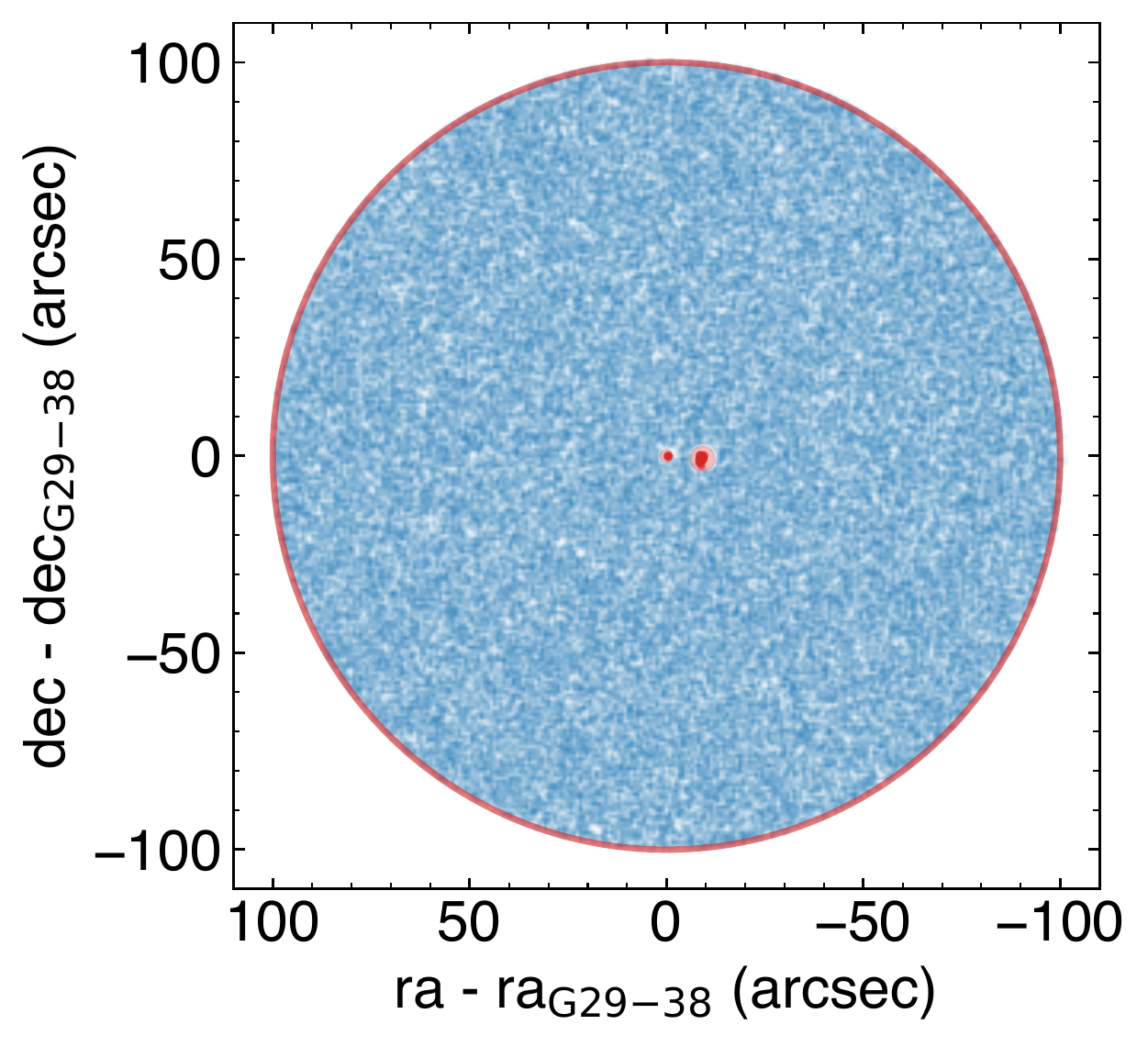}
    \includegraphics[angle=0,width=.59\textwidth]{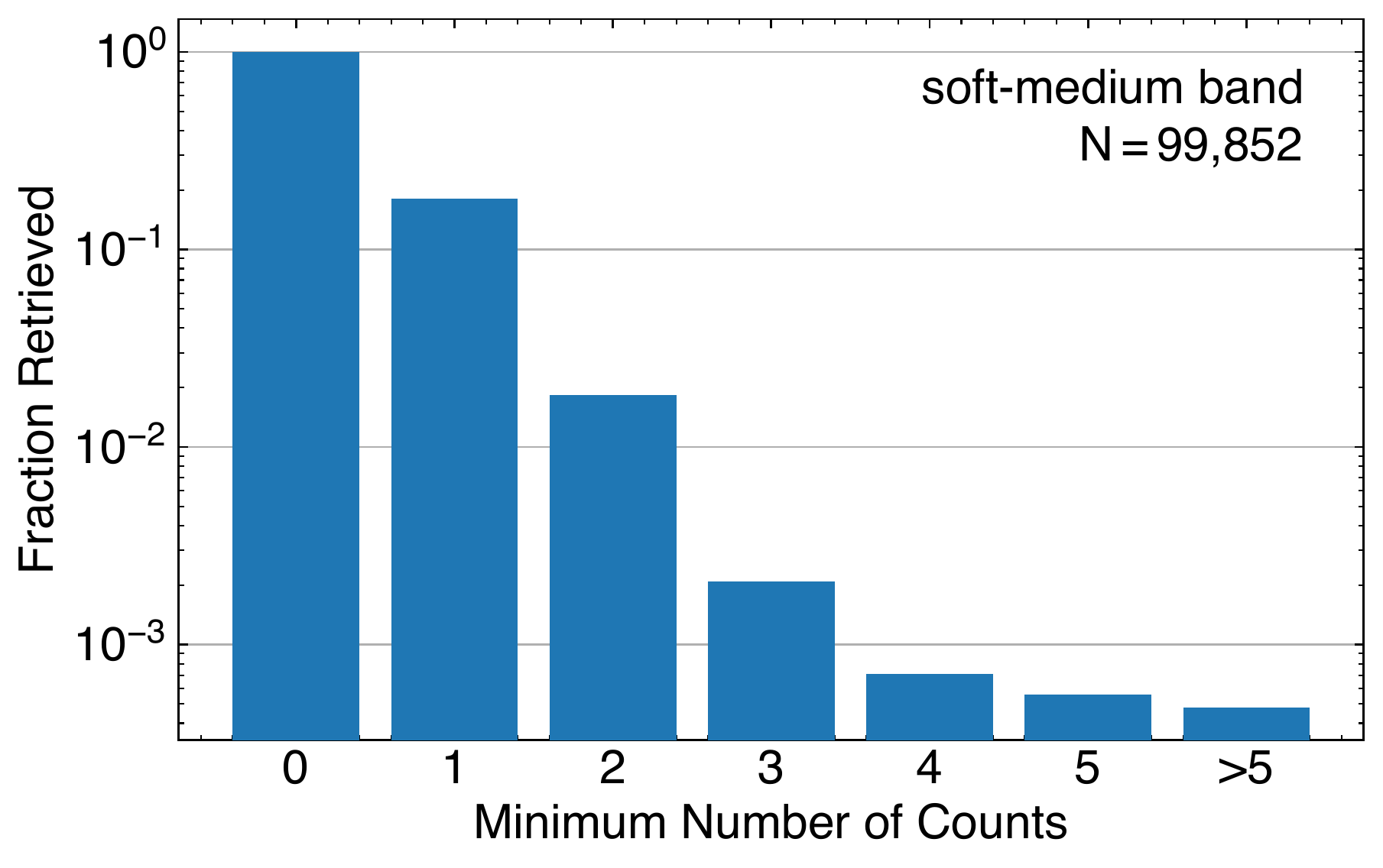}
    \vspace{10pt}
  \caption{\textbf{Monte Carlo aperture photometry.} 
  \textbf{a} \& \textbf{c}, The blue points show the positions of the $\approx$100,000 test apertures, each 1\,arcsec in radius, used to sample 52\,arcsec around the target. The absolute number of test apertures, after removing those that fell within a masked region, is shown in the panels. The sky coordinates of all recorded events that fall within a masked region are shown in orange.  \textbf{b} \& \textbf{d}, The normalised histogram shows the fraction of test apertures with event counts equal to or greater than that of a given bin. The Monte Carlo was performed on the soft (\textbf{a}--\textbf{b}) and soft+medium (\textbf{c}--\textbf{d}) bands. The soft band analysis has 0.001\% of test apertures returning four counts, allowing us to rule out chance alignment at 4.4$\sigma$.
  }
  \label{fg:ap-throw-down-soft-medium-mask}
\end{efigure}

\begin{efigure}
	\centering
	\includegraphics[width=.85\columnwidth]{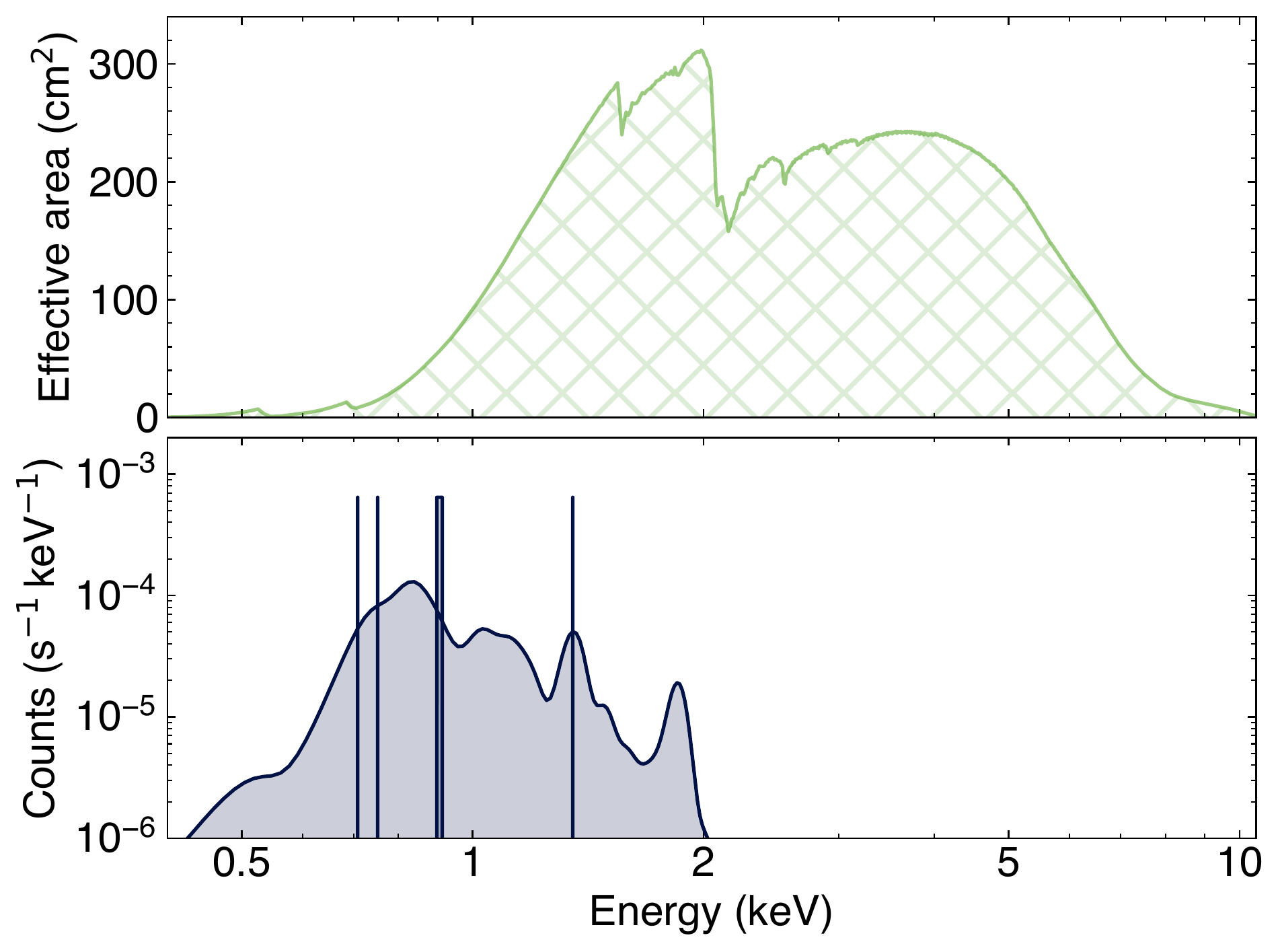}
	\caption{\textbf{Spectral modelling of observed X-ray events}. The observed counts are shown along with the best-fit spectral model for the isothermal plasma with photospheric abundances\cite{xu14} (dark blue). The effective area of the ACIS-S detector is shown in green, hatch. The absence of harder X-ray events ($>$2.0\,keV) in the \textit{Chandra} observations 
	demonstrates that
	the plasma emission spectrum is very soft.}
	\label{fg:counts-effective-area}
\end{efigure}

\begin{efigure}
	\centering
	\includegraphics[width=.7\columnwidth]{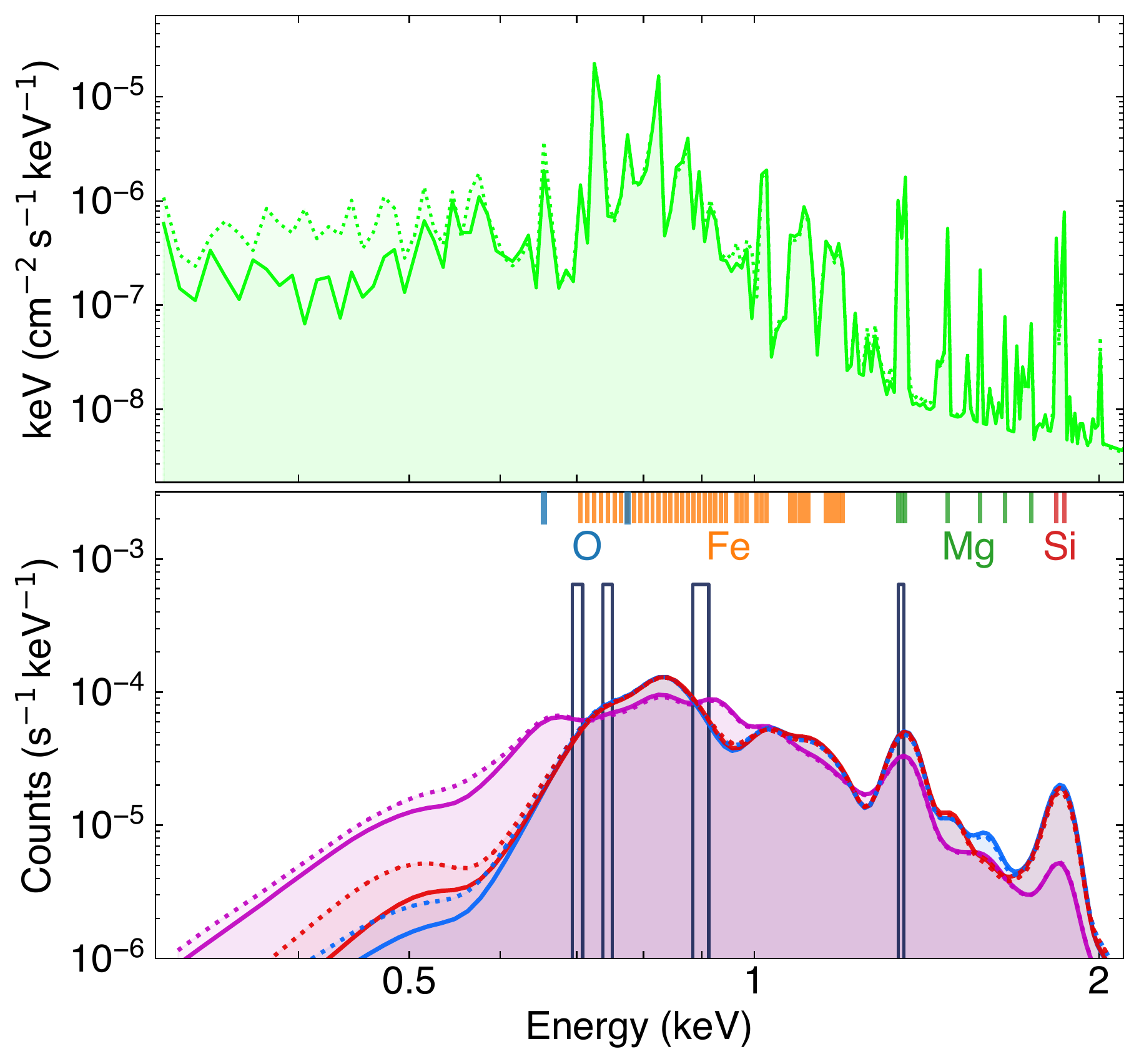}
	\caption{\textbf{Best-fit spectra for the observations using \texttt{XSPEC}}. Bottom, in units of instrumental counts we show the five recorded events (black) and six best-fit spectral models assuming Solar (magenta), bulk Earth\cite{mcdonough1995} (blue) and photospheric\cite{xu14} (red) abundances, with either the \texttt{vvapec} isothermal (solid) and \texttt{mkcflow} cooling flow (dotted) plasma models. We also indicate the dominant metal emission lines (O, Mg, Si and Fe) from the isothermal, photospheric abundance model. Top, in real flux units, we show the synthetic spectra for the photospheric abundances with the isothermal (solid) and cooling flow (dotted) plasma models. The modelling suggests the most likely origin of the source photon at 1.3\,keV was a Mg transition.}
	\label{fg:xspec-fits}
\end{efigure}

\begin{efigure}
	\centering
	\includegraphics[width=.9\columnwidth]{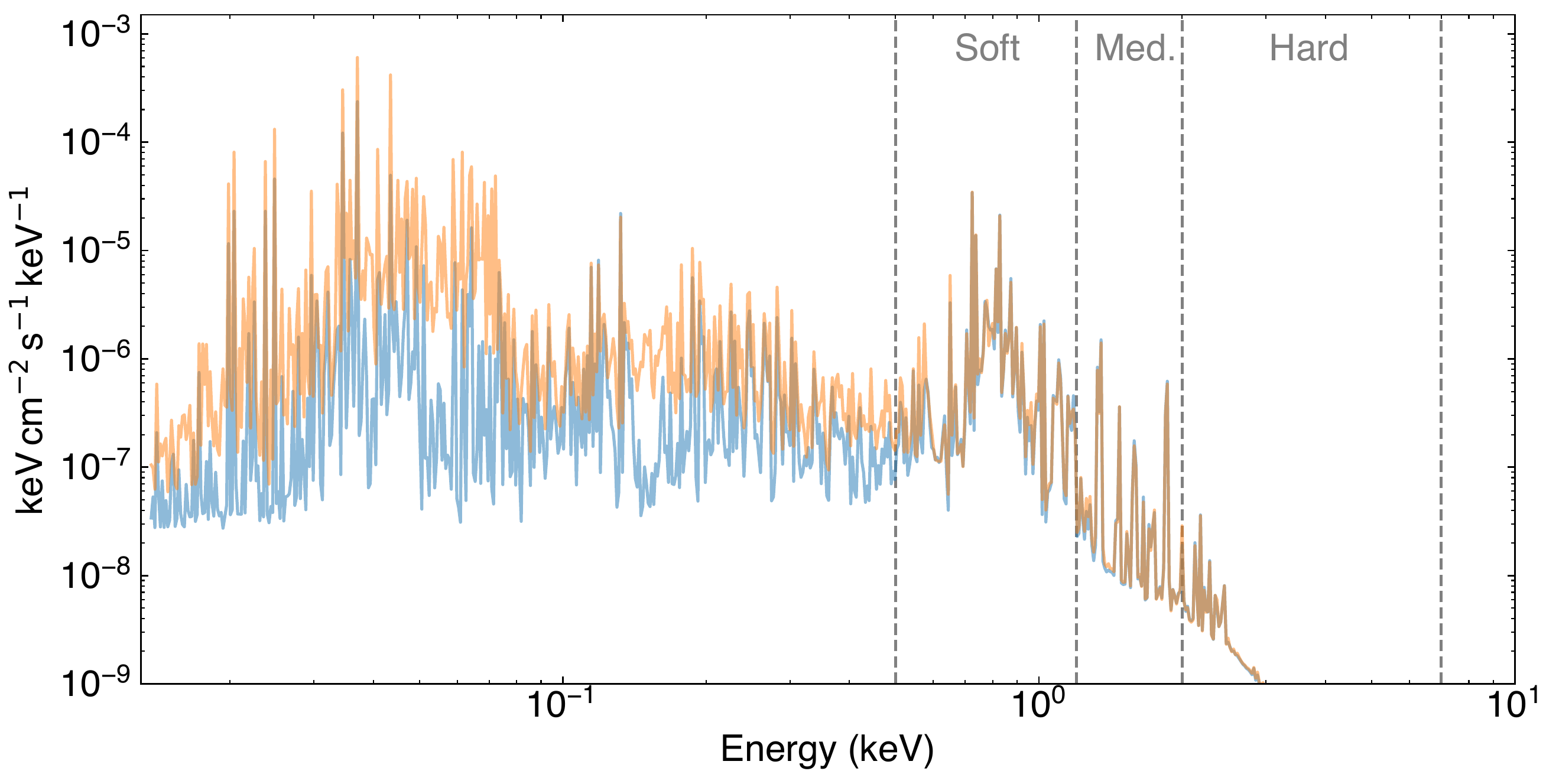}
	\caption{\textbf{Estimate for flux carried at XUV wavelengths}. Spectral energy distribution of the best-fit isothermal (blue) and cooling flow (orange) plasma models with bulk Earth abundances\cite{mcdonough1995} down to the extreme ultraviolet (EUV) energy regime. Also shown are the standard \textit{Chandra} science bands; soft, medium and hard. Both models provide a convergent fit within the \textit{Chandra} ACIS-S passband, but the cooling flow provides a more physical and larger estimate of the lower-energy flux. 
	}
	\label{fg:xuv-spectrum}
\end{efigure}

\clearpage

\begin{efigure}
    \RaggedRight
    \begin{tabularx}{.9\columnwidth}{XXX}
        \textsf{\textbf{a}} & \textsf{\textbf{b}} & \textsf{\textbf{c}}
    \end{tabularx}
    \justifying
	\centering
	\includegraphics[width=1.0\columnwidth]{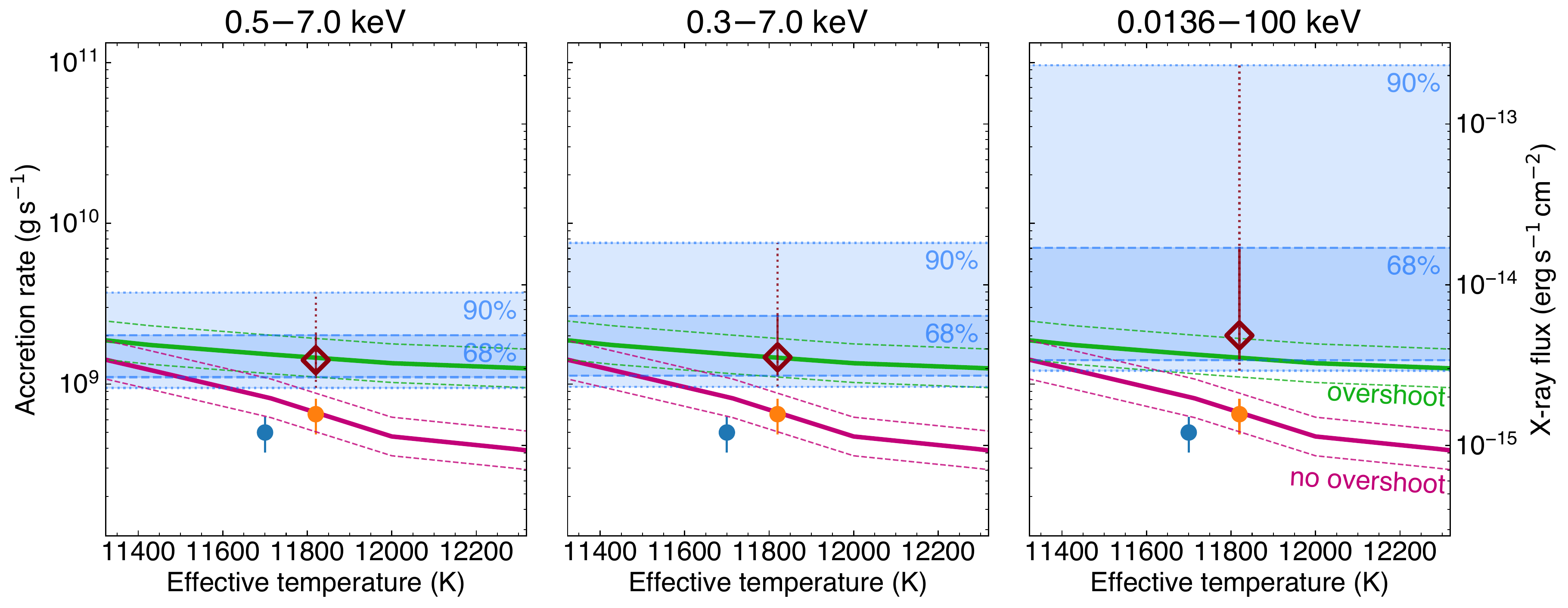}
	\caption{\textbf{X-ray flux and accretion rate.}
	X-ray flux measured in 3 bands: \textbf{(a)} 0.5--7.0\,keV, \textbf{(b)} 0.3--7.0\,keV and \textbf{(c)} 0.0136--100\,keV, using the cooling flow model for the photospheric abundances\cite{xu14} are shown in open diamonds. The filled horizontal bands show the 68\% and 90\% confidence intervals on the X-ray accretion rate, with bounds shown in dashed and dotted lines, respectively. The X-ray accretion rates are computed using Equation\,\eqref{eq:Mdot-Xray}, with $A=1$, $R_{\rm WD}=0.0129R_{\odot}$, and $M_{\rm WD}=0.6M_{\odot}$. 
		The accretion rates inferred from a calcium abundance of $\log{\rm [Ca/H]}=-6.58 \pm0.12$, and assuming bulk Earth composition (such that calcium accounts of 1.6\% of the accreted material), are shown in solid lines. The models including and omitting convective overshoot mixing are shown in green and pink, respectively. Also shown in solid circles (blue and orange) are the previously published inferred accretion rates for G29--38, based on photospheric abundances from spectroscopic observations\cite{farihi09,xu14}.}
	\label{fg:Mdot-1D-3D-mkcflow-3panel}
\end{efigure}

\begin{efigure}
    \RaggedRight
    \begin{tabularx}{\columnwidth}{XXXXXXXXXXXXXXXX}
        \textsf{\textbf{a}} & &&&&&& \textsf{\textbf{b}} &&&&&&&&
    \end{tabularx}
    \centering
    \justifying
	\includegraphics[height=.34\columnwidth]{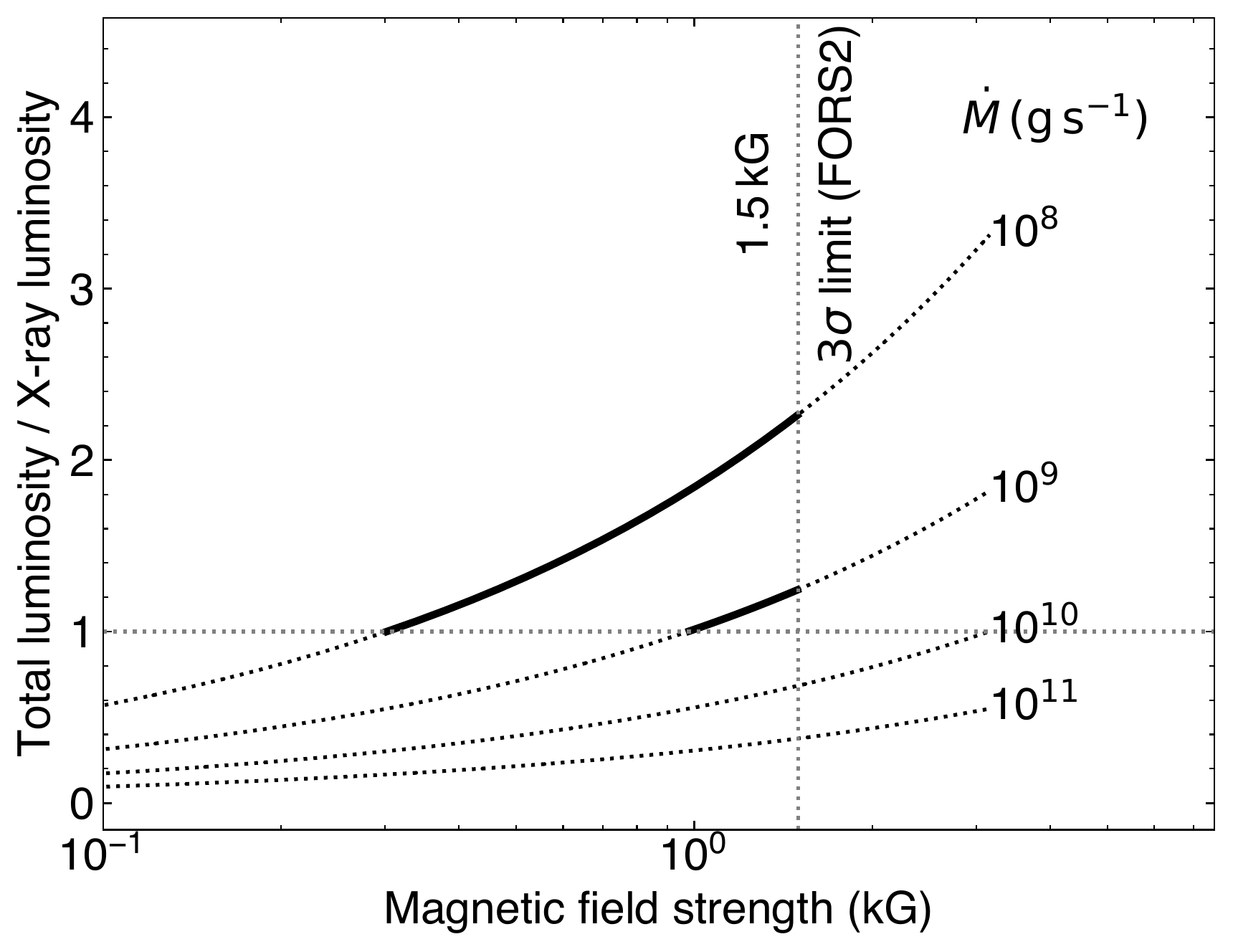} \hfill
	\includegraphics[height=.34\columnwidth]{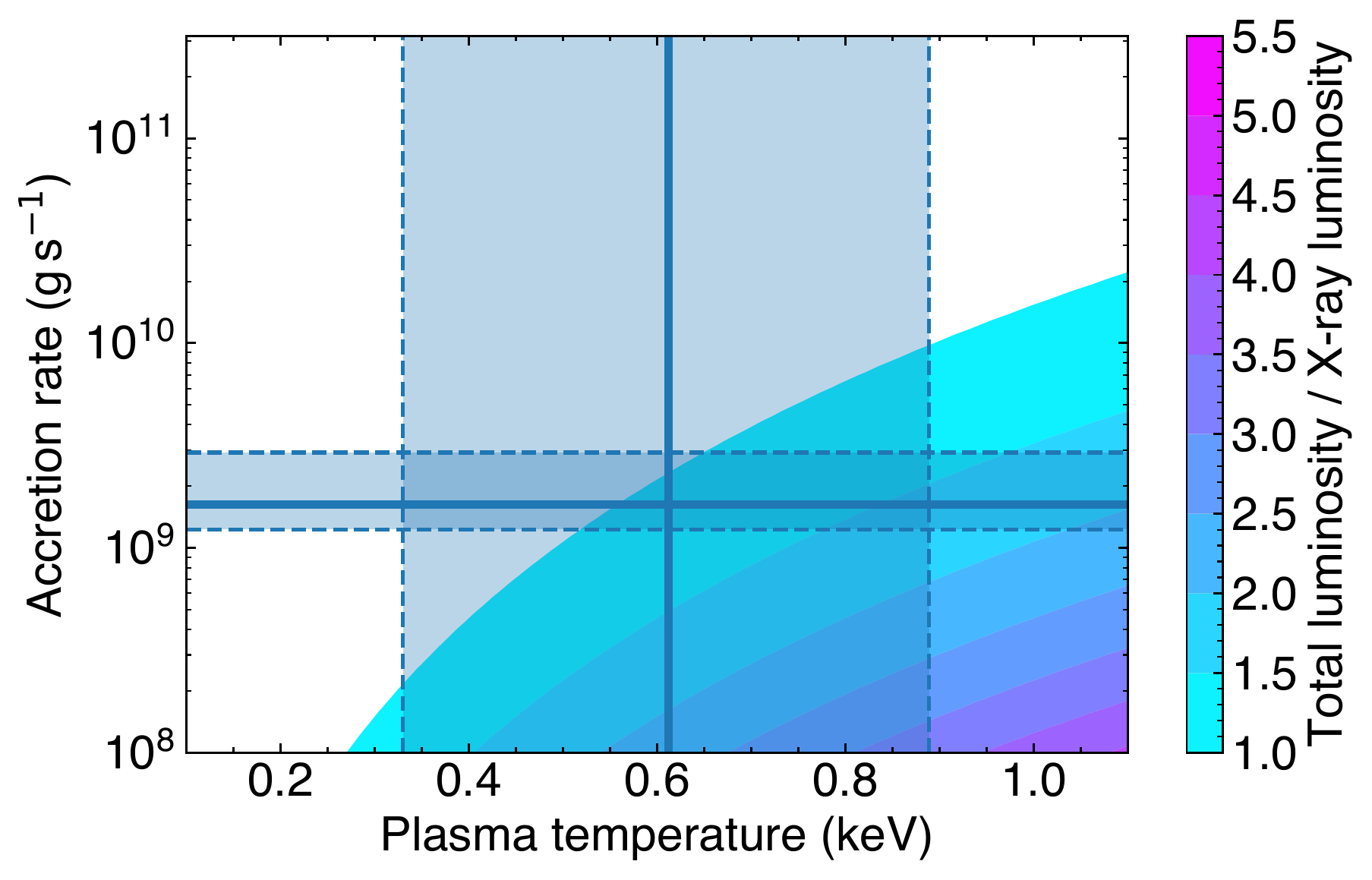}
	\caption{\textbf{Limit on cyclotron emission cooling as source of additional luminosity.} \textbf{a}, An estimate for the total luminosity from the measured X-ray luminosity, accounting for cyclotron emission cooling. We compare the measured plasma temperature, $kT_{\rm X}=(0.61 \pm 0.28)$\,keV, given by the cooling flow model and photospheric abundances, with the critical plasma temperature, $T_B$, above which cyclotron emission cooling dominates, defined by equation 10 from ref.\cite{farihi2018XRay}.
	The authors provide the ratio $T_{\rm X}/T_B\approx L_{\rm tot}/L_{\rm X}$  as an estimate of, for a range of accretion rates and global magnetic field strengths, the predicted increase in total luminosity compared to X-ray luminosity if the plasma temperature is sufficient to be dominated by cyclotron emission cooling. The horizontal dotted line indicates $T_{\rm X}/T_B=1$, where no correction is expected below this. The vertical dotted line indicates the 3$\sigma$ upper limit on the magnetic field strength from FORS2 spectropolarimetric observations\cite{farihi2018XRay}. The solid lines indicate the increase in total luminosity when compared to the observed X-ray luminosity. \textbf{b}, Predicted additional luminosity for an assumed global magnetic field at the 3$\sigma$ limit (1.5\,kG) across the full range of plasma temperatures and accretion rates calculated in this work (see Extended Data Tables\,\ref{tab:confidence-interval-kT}\,\&\,\ref{tab:confidence-interval-Mdot}). White space indicates no additional luminosity. The upper 
	plasma temperature from the cooling flow model and accretion rate derived from the isothermal plasma model is shown (solid) along with the 68\% uncertainty (dashed). Even at the observational upper limit, the predicted increase due to cyclotron emission cooling is a factor of 
	$1.1^{+0.6}_{-0.1}$.}
	\label{fg:lum-increase-cyclotron}
\end{efigure}

\begin{etable}
    \centering
        \caption{\label{tab:Chandra-observations}\textbf{\textit{Chandra} observations of G29--38.}}
    \begin{tabular}{c c c c c} 
    \hline \hline
    Obs ID & Instrument & Exposure Time & Start Date & Epoch \\[-0.5ex]
        & & (ks)         & (yyyy-mm-dd) & (yr) \\ [0.5ex]
    \hline
    24257 & ACIS-S & 24.58 & 2020-09-22 & J2020.727 \\ [-1.6ex]
    24256 & ACIS-S & 14.89 & 2020-09-24 & J2020.732 \\ [-1.6ex]
    24658 & ACIS-S & 14.89 & 2020-09-24 & J2020.732 \\ [-1.6ex]
    23379 & ACIS-S & 26.23& 2020-09-26 & J2020.738 \\ [-1.6ex]
    24657 & ACIS-S & 25.74 & 2020-09-27 & J2020.740 \\ [-.9ex]
    Total: & -- & 106.33 & -- & -- \\ [1ex] 
    \hline
    \end{tabular}
    \\
    \justifying
    \vspace{15pt}
    \noindent
    Details of \textit{Chandra} observations carried out between 22 and 27 September 2020 with G29--28 as the target (PI: Cunningham).
\end{etable}

\begin{etable}
    \centering
        \caption{\label{tab:Poisson-source-and-background} \textbf{Statistical significance of source detection.}}
    \begin{tabular}{l c c c c c c c c} 
    \hline \hline
    Band & Energy & Source & \multicolumn{2}{c}{Background}
    & \multicolumn{2}{c}{$\mathrm{CL}$} & \multicolumn{2}{c}{Significance} \\[-0.5ex]
         & (keV)  &   (per ap.)  & \multicolumn{2}{c}{(per ap.)}
         & \multicolumn{2}{c}{(\%)} & \multicolumn{2}{c}{($\sigma$)}\\ [-0.5ex]
        &   &   & $r_{1.0}$ & $r_{0.5}$ &
         $r_{1.0}$ & $r_{0.5}$ & $r_{1.0}$ & $r_{0.5}$\\ [0.5ex]
    \hline
    Soft & 0.5--1.2 & 4  & 0.100 & 0.025 & 99.99962 & 99.9999984 & 4.62 & 5.65\\ [-1.6ex]
    Soft+Medium & 0.5--2.0 & 5  & 0.205 & 0.051 & 99.99975 & 99.9999997 & 4.71 & 5.94\\ [-1.6ex]
    Broad & 0.5--7.0 & 5  & 0.566 & 0.142 & 99.97255 & 99.9999590 & 3.64 & 5.06\\ [1ex]
    \hline
    \end{tabular}
    \\
    \justifying
    \vspace{15pt}
    \noindent
    Source counts ($N$) and expected background ($b$) in the three standard ACIS energy bands used in this study, given in units of counts per 1\,arcsec aperture ($r_{1.0}$). From top to bottom, the total background counts in each band was $N_b$=270, 555 and 1530 (in the 52\,arcsec radius background region). The expected background in the 1\,arcsec radius source region was thus computed as $b=N_b/52^2$. This includes reprocessing using the \texttt{VFAINT} mode for background cleaning. For an expected background, $b$, the Poisson distribution gives the probability of receiving $N$ or more counts as $P(N | b)= {b^N e^{-b}}/(N!)$. We therefore reject the null hypothesis of detecting no source photons with a confidence of $\mathrm{CL}=100\times(1-P(N|b))$. We also provide results for the relocalised 0.5\,arcsec aperture ($r_{0.5}$), and note that the source counts are the same for both aperture sizes.
\end{etable}

\begin{etable}
	\centering
		\caption{\label{tab:wavdetect-results-table}\textbf{Sky density of sources from the \texttt{wavdetect} source detection algorithm.}}
	\begin{tabular}{l c c c c c c c c c c c} 
		\hline \hline
		Band  & & \multicolumn{2}{c}{No. Sources}  & & \multicolumn{2}{c}{Target?}  & & \multicolumn{2}{c}{Sky Density} & \multicolumn{1}{c}{$\mathrm{CL_{align}}$} & Significance\\[-1.5ex]
		&    & &     & & & & &\multicolumn{2}{c}{($10^{-5}$\,arcsec$^{-2}$)} &\multicolumn{1}{c}{(\%)} & ($\sigma$)\\ [-1.5ex]
		&   & $s_1$& $s_2$ & & $s_1$& $s_2$  & & $s_1$ & $s_2$ & $s_1$ & $s_1$\\ [0.5ex]
		\hline
		Soft &  & 7  & 6 &  & \checkmark & $\times$  & & 2.76 & 2.36 & 99.99724 & 4.2\\ [-1.6ex]
		Soft+Medium &  & 25  & 23 &  & \checkmark & \checkmark  && 9.85 & 9.06 & 99.99015 & 3.9\\ [-1.6ex]
		Broad &  & 32  & 32 &  & $\times$ & $\times$  && 12.6 & 12.6 & 99.98740 & 3.8\\ [1ex] 
		\hline
	\end{tabular}
    \\
    \justifying
    \vspace{15pt}
    \noindent	
	We present results in each of the three standard science bands considered (soft, soft+medium and broad), for two values of the significance threshold parameter, \texttt{sigthresh}, where $s_1$ and $s_2$ are significance thresholds of $1\times 10^{-6}$ and $5\times 10^{-7}$. From the documentation, the significance threshold should be set to $s=1/n_{\rm px}$, which typically produces one false detection. The ACIS-S S3 CCD has 1024$\times$1024 pixels, so $s_1$ should be a sufficiently low threshold. However, to align with the world coordinate system (WCS), the CCD image is rotated relative to the bounding box of the image, meaning that the image on which \texttt{wavdetect} was run has 1414$\times$1401 pixels, hence the value of $s_2=1/(1414\times1401)\approx5\times 10^{-7}$. The sky density is computed as the number of sources, divided by the field of view, which is estimated as ($1024^{2}\times0.4920^{2}$), where the second number is the pixel size in arcsec. 
	Interpreting the sky density as the probability of chance alignment, we provide the confidence, $\mathrm{CL_{align}}$, with which we can reject the hypothesis that our source counts could originate from a background source.
\end{etable}

\begin{etable}
    \centering
        \caption{\label{tab:confidence-interval-counts}\textbf{Confidence interval on \textit{Chandra} ACIS-S count rate.}}
    \begin{tabular}{l c c c c c c c} 
    \hline \hline
        Band & Energy & \multicolumn{6}{c}{Count rate (ap.$^{-1}$\,ks$^{-1}$)} \\[-0.5ex]
     &  & \multirow{2}{*}[0.8ex]{Source}  & \multirow{2}{*}[0.8ex]{Background} & \multicolumn{2}{c}{68\%} & \multicolumn{2}{c}{90\%}\\[-0.5ex]
         & (keV)  &  &  & low & high & low & high \\ [0.5ex]
    \hline
    Soft & 0.5--1.2 & 0.038 & 0.0010 & 0.021 & 0.059 & 0.013 & 0.078 \\ [-1.6ex]
    Soft+Medium & 0.5--2.0 & 0.047 & 0.0021 & 0.027 & 0.070 & 0.018 & 0.090 \\ [-1.6ex]
    Broad & 0.5--7.0  & 0.047 & 0.0057 & 0.024 & 0.066 & 0.015 & 0.086 \\ [1ex]
    \hline
    \end{tabular}
    \\
    \justifying
    \vspace{15pt}
    \noindent
     The 68\% and 90\% confidence intervals for the source counts ($N$) given in Extended Data Table\,\ref{tab:Poisson-source-and-background}, which have been converted to a count rate using the 106.33\,ks total exposure time for the three standard \chandra\ energy bands used in this study. Confidence intervals are calculated using the method of ref.\cite{kraft1991}, which is a Bayesian approach to Poisson statistics in the presence of a background that uses a simple prior on the number of source counts not being negative. All count rate values are given in units of counts per ks, per 1\,arcsec aperture.
\end{etable}

\begin{etable}
    \centering
        \caption{\label{tab:confidence-interval-kT}\textbf{Plasma temperature from spectral modelling using \texttt{XSPEC}.}}
    \begin{tabular}{l l c c c c c} 
    \hline \hline
    \multirow{2}{*}{Abundance} &  \multirow{2}{*}{Model} & \multicolumn{5}{c}{Plasma Temperature (keV)}\\[-1.5ex]
        &   & \multirow{2}{*}[0.8ex]{Best-fit} & \multicolumn{2}{c}{68\%} & \multicolumn{2}{c}{90\%} \\ [-1.5ex]
        &  &  & low & high & low & high \\ [0.5ex]
    \hline
    \multirow{2}{*}[0.8ex]{Solar\cite{CIAO2006fornaturesubmission}} & Isothermal & 0.273 & 0.198 & 0.401 & 0.165 & 0.653 \\ [-1.6ex]
     & Cooling flow  & \textbf{0.317} & 0.226 & 0.581 & 0.187 & 0.922 \\ [-.6ex]
    \multirow{2}{*}[0.8ex]{Bulk Earth\cite{mcdonough1995}} & Isothermal & 0.473 & 0.231 & 0.653 & 0.172 & 0.772 \\ [-1.6ex]
     & Cooling flow & \textbf{0.594} & 0.273 & 0.882 & 0.193 & 1.107 \\ [-.6ex]    
    \multirow{2}{*}[0.8ex]{Photospheric\cite{xu14}} & Isothermal & 0.491 & 0.283 & 0.661 & 0.185 & 0.778 \\ [-1.6ex]
     & Cooling flow & \textbf{0.612} & 0.330 & 0.889 & 0.210 & 1.111 \\ [1ex] 
    \hline
    \end{tabular}
    \\
    \justifying
    \vspace{15pt}
    \noindent
    The 68\% and 90\% confidence intervals for the derived plasma temperature are based on fits to the full ACIS-S spectrum (plotted in Extended Data Figure \ref{fg:xspec-fits}). Best fits and confidence intervals were determined using the C-statistic,\cite{cash79} and with no background subtraction. For the cooling flow model these are the upper temperature of a range of temperatures extending down to 0.08\,keV, with the emission measure at each temperature weighted by the inverse of the emissivity.
\end{etable}

\begin{etable}
    \centering
        \caption{\label{tab:confidence-interval-Flux}\textbf{Best-fit X-ray flux from spectral models.}}
    \begin{tabular}{l c c c c c} 
    \hline \hline
    \multirow{2}{*}{Energy} &  \multicolumn{5}{c}{Flux ($10^{-15}\,\mathrm{erg\,s^{-1}\,cm^{-2}}$)}\\[-0.5ex]
     &  \multirow{2}{*}[0.8ex]{Best-fit} & \multicolumn{2}{c}{68\%} & \multicolumn{2}{c}{90\%} \\ [-1.5ex]    
        (keV) &  & low & high & low & high \\ [0.5ex]
    \hline
    0.5--7.0    & \textbf{1.78}  & 1.40  & 2.53  & 1.20  & 4.53  \\ [-1.6ex]
    0.3--7.0    & \textbf{1.97}  & 1.49  & 3.52  & 1.26  & 9.40  \\ [-1.6ex]
    0.0136--100 & \textbf{4.34}  & 2.62  & 19.1 & 2.09  & 218.0 \\ [1ex]
    \hline
    \end{tabular}
    \\
    \justifying
    \vspace{15pt}
    \noindent
    The best-fit fluxes are computed across three spectral energy ranges using the cooling flow model with photospheric abundances\cite{xu14}. The first two energy ranges are within the wavelength range of our observations and provide a robust lower limit on the flux. The third, wider energy band allows for more flux to be carried at higher and lower energies, thus providing an upper estimate of the X-ray flux based on our observations. The 68\% and 90\% confidence intervals for the derived flux based on those for the plasma temperature in Extended Data Extended Data Table\,\ref{tab:confidence-interval-kT}.
\end{etable}

\clearpage

\begin{etable}
    \centering
    \caption{\label{tab:confidence-interval-luminosity}\textbf{X-ray luminosity of G29--38.}}
    \begin{tabular}{l c c c c c} 
    \hline \hline
    \multirow{2}{*}{Energy} &  \multicolumn{5}{c}{X-ray Luminosity ($10^{25}\,\mathrm{erg\,s^{-1}}$)}\\[-0.5ex]
     &  \multirow{2}{*}[0.8ex]{Best-fit} & \multicolumn{2}{c}{68\%} & \multicolumn{2}{c}{90\%} \\ [-1.5ex]    
        (keV) &  & low & high & low & high \\ [0.5ex]
    \hline
    0.5--7.0    & \textbf{6.54}  & 5.15 &	9.30 &	4.41 &	16.7  \\ [-1.6ex]
    0.3--7.0    & \textbf{7.24}  & 5.48 &	12.9  &  4.63 &	34.5  \\ [-1.6ex]
    0.0136--100 & \textbf{16.0} & 9.63	& 70.2 &  7.68 & 801.3 \\ [1ex] 
    \hline
    \end{tabular}
    \\
    \justifying
    \vspace{15pt}
    \noindent  
    The best-fit X-ray luminosity ($L_{\rm X}$) derived using $L_{\rm X}=4\pi d^{2} F_{\rm X}$ for the fluxes ($F_{\rm X}$) given in Extended Data Table\,\ref{tab:confidence-interval-Flux}.
\end{etable}

\clearpage

\begin{etable}
    \centering
    
    \caption{\label{tab:confidence-interval-Mdot}\textbf{X-ray accretion rate of G29--38.}}
    \begin{tabular}{l c c c c c} 
    \hline \hline
    \multirow{2}{*}{Energy} &  \multicolumn{5}{c}{X-ray Accretion Rate ($10^{9}\,\mathrm{g\,s^{-1}}$)}\\[-0.5ex]
     &  \multirow{2}{*}[0.8ex]{Best-fit} & \multicolumn{2}{c}{68\%} & \multicolumn{2}{c}{90\%} \\ [-1.5ex]    
        (keV) &  & low & high & low & high \\ [0.5ex]
    \hline
    0.5--7.0    & \textbf{1.48}   & 1.16 &	2.09 &	1.00 &	3.76    \\ [-1.6ex]
    0.3--7.0    & \textbf{1.63}   & 1.23 &	2.92 &	1.05 &	7.80    \\ [-1.6ex]
    0.0136--100 & \textbf{3.60}   & 2.17 &	15.8 &	1.73 &	181 \\ [1ex] 
    \hline
    \end{tabular}
    \\
    \justifying
    \vspace{15pt}
    \noindent  
    The best-fit X-ray accretion rate ($\dot{M}_{\rm X}$) derived using $\dot{M}_{\rm X}=2R_{\rm WD}/ (G M_{\rm WD})$ with the luminosities given in Extended Data Table\,\ref{tab:confidence-interval-luminosity}.
\end{etable}


\end{document}